\documentclass[journal,final]{ieeetj}

\usepackage{float}
\usepackage{amsbsy}
\usepackage{amssymb}
\usepackage{amsmath}
\usepackage[math]{cellspace}
\usepackage{textcomp}

\usepackage{soul, xcolor}

\usepackage{cite,enumerate}
\usepackage{array} 
\usepackage{adjustbox}
\usepackage{makecell}
\usepackage{setspace}

\def\BibTeX{{\rm B\kern-.05em{\sc i\kern-.025em b}\kern-.08em
    T\kern-.1667em\lower.7ex\hbox{E}\kern-.125emX}}
\AtBeginDocument{\definecolor{tmlcncolor}{cmyk}{0.93,0.59,0.15,0.02}\definecolor{NavyBlue}{RGB}{0,86,125}}

\def\OJlogo{\vspace{-4pt}$<$Society logo(s) and publication title will appear here.$>$}
\def\seclogo{\vspace{10pt}$<$Society logo(s) and publication title will appear here.$>$}

\def\OJlogo{} 
\def\seclogo{} 

\def\authorrefmark#1{\ensuremath{^{\textbf{#1}}}}

\usepackage{tikz}
\usetikzlibrary{shapes}
\usetikzlibrary{arrows}
\usetikzlibrary{patterns}
\usetikzlibrary{arrows.meta}
\usetikzlibrary{decorations} 
\usetikzlibrary{decorations.pathreplacing}
\usetikzlibrary{decorations.pathmorphing}
\usetikzlibrary{shapes,backgrounds,patterns,decorations.text}
\usetikzlibrary{decorations.markings}
 
\usepackage{mathtools}
\mathtoolsset{showonlyrefs}

\def\O{{{\mathcal{O}}}}

\def\bA{{\boldsymbol{A}}}

\def\bE{{\boldsymbol{E}}}

\def\bI{{\boldsymbol{I}}}

\def\bL{{\boldsymbol{L}}}

\def\bQ{{\boldsymbol{Q}}}

\def\bS{{\boldsymbol{S}}}

\def\bW{{\boldsymbol{W}}}
\def\bX{{\boldsymbol{X}}}
\def\bY{{\boldsymbol{Y}}}
\def\bZ{{\boldsymbol{Z}}}

\def\ba{{\boldsymbol{a}}}

\def\bb{{\boldsymbol{b}}}

\def\bcO{{\boldsymbol{\mathcal{O}}}}
\def\bd{{\boldsymbol{d}}}

\def\beps{{\boldsymbol{\epsilon}}}

\def\be{{\boldsymbol{e}}}

\def\bg{{\boldsymbol{g}}}

\def\bone{{\boldsymbol{1}}}

\def\bq{{\boldsymbol{q}}}

\def\bs{{\boldsymbol{s}}}

\def\bv{{\boldsymbol{v}}}
\def\bu{{\boldsymbol{u}}}

\def\bx{{\boldsymbol{x}}}
\def\by{{\boldsymbol{y}}}
\def\bzero{{\boldsymbol{0}}}
\def\bz{{\boldsymbol{z}}}

\def\realR{{\mathbb{R}}}

\def\eqa{\, { \mathop = \limits^{\scriptsize{\mbox{(a)}}}} \,}
\def\eqb{\,{ \mathop = \limits^{\scriptsize{\mbox{(b)}}}} \,}

\def\lea{\,{ \mathop \le \limits^{\scriptsize{\mbox{(a)}}}}\,}
\def\leb{\,{ \mathop \le \limits^{\scriptsize{\mbox{(b)}}}}\,}
\def\lec{\,{ \mathop \le \limits^{\scriptsize{\mbox{(c)}}}}\,}

\def\naf{{{\nabla f(\bx)}}}
\def\nafs{{{\nabla f(\bx_*)}}}

\def\bxl{{{\bx^*_{\scriptsize\mbox{loc}}}}}
\def\bxhs{{{\hat\bx-\bx_*}}}

\def\Piu{{{\Pi_\bone}}}
\def\Piup{{{\Pi_\bone^\perp}}}
\def\bIW{{{(\bI-\bW)}}}
\def\bIWp{{{(\bI-\bW)^\dagger}}}

\def\kmax{{k_{\scriptsize\mbox{max}}}}

\newcommand{\pp}[1]{{\left( #1 \right)}}
\newcommand{\ppb}[1]{{\left[ #1 \right]}}
\newcommand{\norm}[1]{{ \left\Vert #1 \right\Vert }}
\newcommand{\snorm}[1]{{ \left\Vert #1 \right\Vert^2 }}
\newcommand{\mnorm}[1]{{ \left\Vert #1 \right\Vert_2 }}

\newcommand{\E}[1]{{ \mathsf{E} \{ #1  \} }}
\newcommand{\Ebig}[1]{{ \mathsf{E} \left\{ #1  \right\} }}

\newcommand{\cov}[1]{{ \mathsf{Cov} \{ #1  \} }}
\newcommand{\var}[1]{{ \mathsf{Var} \{ #1  \} }}

\newcommand{\tr}[1]{{ \mathsf{Tr} \left\{ #1  \right\} }}
\renewcommand{\vec}[1]{{ \mathsf{vec} \left\{ #1  \right\} }}

\def\T{{{\text{T}}}}

\DeclareMathOperator*{\argmin}{argmin}
 
\title{Unified Analysis of Decentralized Gradient Descent: \\ a Contraction Mapping Framework}

\author{Erik G. Larsson\authorrefmark{1} and Nicolò Michelusi\authorrefmark{2}}
\affil{Link\"oping University, Dept. of Electrical Engineering (ISY), 581 83 Link\"oping, Sweden.}
\affil{School of Electrical, Computer and Energy Engineering, Arizona State University, United States} 
\authornote{Erik G. Larsson's research is funded by the KAW
  Foundation, ELLIIT, and the Swedish Research Council (VR).  \\ Nicolò
  Michelusi's research is funded in part by NSF under grant
  CNS-$2129015$.}

\begin{document}

\currentdate{\today} 
\doiinfo{10.1109/OJSP.2025.3557332} 
	
\begin{abstract}
 The decentralized gradient descent (DGD) algorithm, and its sibling,
 diffusion, are workhorses in decentralized machine learning,
 distributed inference and estimation, and multi-agent coordination.
 We propose a novel, principled framework for the analysis of DGD and
 diffusion for strongly convex, smooth objectives, and arbitrary
 undirected topologies, using contraction mappings coupled with a
 result called the mean Hessian theorem (MHT).  The use of these tools
 yields tight convergence bounds, both in the noise-free and noisy
 regimes.  While these bounds are qualitatively similar to results
 found in the literature, our approach using contractions together
 with the MHT decouples the algorithm dynamics (how quickly the
 algorithm converges to its fixed point) from its asymptotic
 convergence properties (how far the fixed point is from the global
 optimum). This yields a simple, intuitive analysis that is accessible
 to a broader audience.  Extensions are provided to multiple local
 gradient updates, time-varying step sizes, noisy gradients
 (stochastic DGD and diffusion), communication noise, and random
 topologies.
\end{abstract}


\maketitle

\section{Introduction}\label{sec:intro}

In applications such as decentralized machine learning
\cite{YANG2019278}, distributed inference and estimation
\cite{9224135,6494683}, and multi-agent coordination \cite{Nedic2018},
$N$ agents interconnected in a communication network cooperate on the
minimization of a global objective function ${F(x)=\sum_{n=1}^N
  f_n(x)/N}$, where $f_n(x)$ represents the local objective function
at the $n$th agent.  The algorithmic workhorse for minimizing $F(x)$
is decentralized gradient descent (DGD), and its noisy (stochastic)
versions.

This paper presents a textbook-style, from-first-principles analysis
of DGD and its variants for strongly convex, smooth objectives, and
arbitrary, undirected network topologies.  Our focus on strongly
convex and smooth optimization problems is motivated by their stronger
convergence guarantees, simplified analysis, and theoretical insights.
In fact, the insights gained from their analysis can often guide the
design of novel algorithms in more complex settings, serving as a
cornerstone for both theoretical and practical developments.

Our main analytical tools are the mean Hessian theorem (MHT), and a
succession of theorems that deal with contraction mappings: the
contraction convergence theorem (C$^2$T), the concatenated C$^2$T
(C$^3$T), and the noisy C$^3$T (NC$^3$T).  We first use these
contraction theorems together with the MHT to establish a tight
contraction to a fixed point, and then, independently, use the MHT to
find a tight bound on the distance from this fixed point to the
optimal solution.  Via the MHT, the difference between the gradients
at different points $\bx$ and $\by$ can be expressed as a linear
function of ${\by-\bx}$, akin to a quadratic function: ${\nabla f(\by)
  - \naf = \bA_{\bx\by}(\by-\bx)}$, where $\bA_{\bx\by}$ is a positive
definite matrix (the mean Hessian).  The MHT allows leveraging
properties of positive definite matrices in the analysis of DGD
algorithms, significantly simplifying their analysis. In particular,
we show that the MHT encapsulates all key properties of strong
convexity and smoothness, which are typically used separately in
convergence proofs, within a single linear equation.

Beyond the basic DGD algorithm, we cover a number of extensions,
including: the diffusion version of DGD, multiple local gradient
updates, time-varying step sizes, sampling noise (stochastic DGD),
communication noise, random topologies, and link failures.   

\subsection{Literature on DGD and its Variants}\label{subsec:lit}

The core ideas behind the DGD algorithm go back at least to
\cite{tsitsiklis1986distributed}.  The literature on its analysis is
quite large. The analysis in \cite{Nedic2009grad} uses an approach
with subgradients to handle objectives that are convex but not
necessarily smooth, and connectivity graphs that vary over time.  A
key assumption is that the subgradients are uniformly bounded and that
the time-variation of the connectivity satisfies certain conditions.
In \cite{yuan2016convergence}, an analysis is given for smooth and
convex, but not necessarily strongly convex, functions.  Several
variations of the basic DGD algorithm have been developed.  An
important example is the diffusion algorithm \cite{8491372}.  Other
extensions of DGD are gradient tracking
\cite{doi:10.1137/16M1084316,sun2022distributed,7398129} and DGD for
optimization over directed topologies \cite{xin2020general}.

A special case is when the communication network topology is a
complete graph. In this case, popularized as \emph{federated learning}
in \cite{mcmahan2017communication}, in each iteration every agent
exchanges information with everyone else, typically via a central
server.\footnote{\setstretch{0.7}The interconnection topology is a
star (with the server at the center), but since the only role of the
server is to perform aggregation, the effective topology for the
decentralized optimization becomes a complete graph.}  DGD (or
diffusion) for the complete-graph scenario is sometimes called
\emph{distributed} gradient descent, and its analysis follows as a
special case.

Our exposition targets a broad audience, providing an accessible
exposition of the main principles and an entry point to the field. The
uniqueness of our treatment lies in the MHT+contraction framework that
allows a separate analysis of the convergence to a fixed point, and
the distance from this fixed point to the optimal point.  Among other
papers with unified analyses of decentralized optimization algorithms
we note especially
\cite{xu2021distributed,koloskova2020unified,wang2021cooperative} --
these papers contain a comprehensive treatment of the convergence rate
for many different algorithms under a rich variety of assumptions,
tailored to a more specialized readership.  Noteworthy are also the
overview papers \cite{nedic2020distributed,xin2020decentralized}, and
the book \cite{sayed2022inference}, with tutorial-style expositions of
different versions of DGD.

In the last few years, a sizeable body of literature has appeared on
federated and fully decentralized learning over random networks
\cite{jakovetic2018convergence}, communication-constrained and
unreliable networks
\cite{8786146,9782148,9562482,wang2022matcha,chiu2023laplacian,mao2023safari},
and wireless networks
\cite{10338845,9014530,8870236,9515709,xing2021federated,9517780,9322286,ye2022decentralized,Michelusi-NCOTADGD,jiang2022decentralized}.
Good overview papers of this literature are
\cite{9579020,10153432,10251949,9562559,gafni2022fed}.  Our
MHT+contraction framework can naturally handle the impact of
communication noise, randomness of the topology, and link failures.

\subsection{Specific Technical Contributions}

Our main novel contribution is the MHT+contraction framework, enabling
a unified analysis of DGD and diffusion. In addition, we make the
following specific contributions:
\begin{enumerate}
  \item The statement and proof of the NC$^3$T theorem for noisy
    contractions.

    While the C$^2$T and C$^3$T are immediate consequences of standard
    fixed point theorems, NC$^3$T is a new and powerful result that
    makes a clean break with previous approaches. The NC$^3$T
    specifically enables the analysis of noisy contraction mappings
    for which the noise variance scales with the argument of the
    mapping.

  \item The statement and proof of the MHT for only \emph{once
  differentiable} functions.

    The MHT for \emph{twice differentiable} functions is well known,
    see the proof of Lemma 1.2.2 in \cite{nesterov2018lectures}, and
    was used before to analyze DGD -- for example in
    \cite{6777576,nassif2020adaptation,chen2012diffusion}.

  \item A discussion of tightness of our convergence bounds for DGD
    and diffusion, showing that the contraction factors are the best
    possible.

  \item A discussion of various special cases, including that of
    bipartite topologies, and of various misconceptions.
\end{enumerate}
 
{The limitations of our framework are: (1) the objectives must be
(once) differentiable, strongly convex, and smooth; and (2) the weight
matrix must be symmetric. Nonetheless, when treating random topologies
(Section~\ref{sec:noisygrad}.\ref{sec:linkfail}), it is sufficient
that the weight matrix is symmetric \emph{in expectation}; individual
realizations of it may be asymmetric and directed.}
 
\subsection{Roadmap}

Section~\ref{sec:tools} gives the main mathematical tools, the
contraction theorems (C$^2$T, C$^3$T, and NC$^3$T), and the MHT.
Section~\ref{sec:GD} shows, as a preparation for the analysis of the
DGD and diffusion algorithms, how the MHT+contraction framework can be
used to analyze gradient descent (GD).  Section~\ref{sec:DGD} contains
the main analysis of DGD and diffusion, for single and multiple local
gradient updates, and for time-varying step sizes.
Section~\ref{sec:noisygrad} extends the analyses to the cases of
gradient sampling noise, communication noise, and random topologies.
Section~\ref{sec:linkfailapp} exemplifies how these analyses apply to
DGD with random link failures.  Section~\ref{sec:discuss} provides
additional discussion and insights, and extensions.
Section~\ref{sec:concl} concludes the paper.

\subsection{Notation and Conventions}

Lowercase bold symbols are column vectors, and uppercase bold symbols
are matrices.  $\bzero$ is the all-zeros vector/matrix. $\bI$ is the
identity matrix, and $\{\be_n\}$ denote the columns of $\bI$.
$(\cdot)^\T$ denotes the transpose of a vector or matrix.  $\bone$ is
the vector of all ones: $\bone=[1,...,1]^\T$.  For a vector $\bx$,
$\Vert\bx\Vert$ is its Euclidean norm.  For a matrix $\bX$, $\bX^{-1}$
is its inverse.  $\vec{\bX}$ denotes the vector obtained by stacking
the columns of a matrix $\bX$ on top of one another, and $\otimes$
denotes the Kronecker product.  ${\Pi_\bX=\bX(\bX^\T\bX)^{-1}\bX^\T}$
denotes the orthogonal projection onto the column space of $\bX$, and
${\Pi^\perp_\bX=\bI-\Pi_\bX}$ denotes the orthogonal projection onto
the orthogonal complement of this space.  $\bX^\dagger$ denotes the
Moore-Penrose pseudoinverse.  For symmetric matrices $\bX$ and $\bY$
of compatible dimensions, $\bX\succcurlyeq\bY$ means that $\bX-\bY$ is
positive semidefinite.  $\mnorm{\bX}$ is the spectral norm of the
matrix $\bX$.  $\nabla$ and $\nabla^2$ denote the gradient and Hessian
operators, respectively; as a special case, $(\cdot)'$ denotes
derivative.  $\O(x)$ stands for big ordo and $\bcO(x)$ denotes a
matrix whose spectral norm is $\O(x)$.  $(\cdot)^+$ stands for
$\max(0,\cdot)$.

Explicit conditions on $\eta$ (to be used for the step size) will be
stated when their dependence on the problem offers significant
insight.  At other places, to avoid digression into uninteresting
calculations, we say that an expression is \emph{valid for small
enough $\eta$} if there exists $\eta'$ such that the expression holds
for $\eta\le\eta'$, without giving $\eta'$ explicitly. For example,
${\eta/\sqrt{2\eta-\eta^2}\le \sqrt{\eta}}$ for small enough $\eta$.

\section{Mathematical Tools}\label{sec:tools}

\subsection{Contractions and Fixed Points}

A function ${\phi:\realR^m \to\realR^m}$ is called a
\emph{contraction} (on $\realR^m$) if there exists a constant $c$,
$c<1$, such that
\begin{align}
\norm{  \phi(\by) - \phi(\bx)  } \le c \norm{ \by-\bx } \label{eq:contraction}
\end{align}
for all $\bx$ and $\by$.  For a contraction $\phi(\cdot)$, Banach's
fixed point theorem \cite[Th.~9.23]{rudin1953principles} states that
there is a unique \emph{fixed point}, $\hat\bx$, such that
\begin{align}
  {\phi(\hat\bx)=\hat \bx}.
\end{align}
 
The following consequence of the fixed point theorem will be one of
our main tools.\\[2mm]
\noindent\textbf{Contraction Convergence Theorem (C$^2$T):} Let
$\phi(\cdot)$ be a contraction as defined in \eqref{eq:contraction},
and consider the recursion
\begin{align}
  \bx_{t+1} = \phi(\bx_t)  \label{eq:FPTseq}
\end{align}
for $t=0,1,2,\ldots$ and some arbitrary $\bx_0$.  Then
\begin{align}
  \norm{ \bx_{t} - \hat\bx} \le c^t \norm{\bx_0 - \hat\bx}.  \label{eq:FPTconv}
\end{align}
In particular, $\bx_t\to\hat\bx$ as $t\to \infty$.$ \hfill\blacksquare$\\

The C$^2$T follows immediately from the fact that
\begin{align}
  \norm{ \bx_{t+1} - \hat\bx} & =   \norm{ \phi(\bx_t) - \phi(\hat\bx)}
  \le c \norm{\bx_t - \hat\bx} , 
\end{align}
together with induction over $t$. \\

A mapping formed by nesting contractions is also a contraction. More
formally we have the following.\\[2mm]
\noindent\textbf{Concatenated Contraction Convergence Theorem
  (C$^3$T):} Let ${\{\phi_1(\cdot),\ldots,\phi_n(\cdot)\}}$ be
contractions, with contraction factors $\{c_1,\ldots,c_n\}$,
respectively.  Then the concatenated mapping
\begin{align}
  \varphi(\bx)=\phi_n(\cdots(\phi_1(\bx))) \label{eq:varphidef1}
\end{align}
is also a contraction.  Moreover, the recursion
\begin{align}
  \bx_{t+1} = \varphi(\bx_t) \label{eq:FPTseqc3}
\end{align}
has the property
\begin{align}
  \norm{ \bx_{t} - \hat\bx} \le \pp{\prod_{i=1}^n c_i}^t \norm{\bx_0 - \hat\bx},
  \label{eq:FPTconvc3}
\end{align}
where $\hat\bx$ is the fixed point of $\varphi(\cdot)$.
$\hfill\blacksquare$\\

\subsubsection*{Proof of the C$^3$T}

Recursive application of \eqref{eq:contraction} shows that
$\varphi(\cdot)$ is a contraction:
\begin{align}
  & \quad  \norm{ \varphi(\by) - \varphi(\bx) } \\
  & = \norm{  \phi_{n}( \cdots(\phi_1(\by)))    - \phi_{n}(\cdots(\phi_1(\bx)))  } \\
  & \leq c_n  \norm{  \phi_{n-1}( \cdots(\phi_1(\by)))    - \phi_{n-1}( \cdots(\phi_1(\bx))) } \\
  & \leq c_nc_{n-1}  \norm{  \phi_{n-2}( \cdots(\phi_1(\by)))    - \phi_{n-2}( \cdots(\phi_1(\bx))) } \\
  & \leq \cdots \le c_nc_{n-1}\cdots c_{2}  \norm{  \phi_{1}( \by)    - \phi_{1}( \bx) } \\
  & \leq  \pp{ \prod_{i=1}^n c_i }  \norm{\by - \bx}.
\end{align}
Banach's fixed point theorem guarantees the existence of $\hat\bx$
such that ${\varphi(\hat\bx)=\hat\bx}$. Then \eqref{eq:FPTconvc3}
follows from the C$^2$T applied to $\varphi(\cdot)$.$
\hfill\blacksquare$
 
 \subsection{Noisy Contractions}\label{sec:noisycontr}

If the recursions in the C$^2$T or C$^3$T contain noise, then under
mild conditions convergence  holds in the mean-square sense. We
give a noisy version of the C$^3$T, from which a noisy version of the
C$^2$T follows as a special case.\\

\noindent\textbf{Noisy C$^3$T (NC$^3$T):} Let
${\{\phi_1(\cdot),\ldots,\phi_n(\cdot)\}}$ be contractions, each with the
same contraction factor $c$. Define a (noisy) mapping from $\bx_t$
onto $\bx_{t+1}$ via the noisy recursions
\begin{align}
  \bx_{t,0} & =\bx_t\\
  \bx_{t,1} & =\phi_1(\bx_{t,0})+\beps_{t,0}\\
  \bx_{t,2} & =\phi_2(\bx_{t,1})+\beps_{t,1}\\
  \cdots\\
  \bx_{t,n} & =\phi_n(\bx_{t,n-1})+\beps_{t,n-1}\\
  \bx_{t+1} & = \bx_{t,n},
\end{align}
where $\{\beps_{t,i}\}$ are noise terms that satisfy
\begin{align}
  \E{ \beps_{t,i} | \bx_{t,i} } & = \bzero,  \label{eq:noise1stmom} \\
  \sqrt{\E{ \snorm{\beps_{t,i}} | \bx_{t,i} }} & \le \sigma + \omega \norm{\bx_{t,i}}, \label{eq:noise2ndmom}
\end{align}
for some non-negative constants $\sigma$ and $\omega$, and suppose
\begin{align}
  c^2+\omega^2 & <1.  \label{eq:comegalt1}
\end{align}
Let the contraction $\varphi(\cdot)$ be defined as in
\eqref{eq:varphidef1}, and let $\hat\bx$ be its fixed point.  Then
\begin{align}
   \hspace*{-5mm}  \sqrt{ \E{ \snorm{\bx_t - \hat\bx}} } 
   \le \hat d + \sqrt{c^2 + \omega^2}^{\,nt} \pp{  {  \norm{\bx_0 - \hat\bx}} - \hat d}^+,
  \label{eq:nc3tmain}
\end{align}
where $\hat d$ and $M$ are constants given by
\begin{align}
  \hat d  &= \frac{\omega M + \sigma}{ \sqrt{1-c^2}-\omega } ,  \label{eq:dhatupbnd} \\
  M & =\max\{ \norm{\tilde\bx_1},\ldots,\norm{\tilde\bx_n} \},
\end{align}
and where $\{ \tilde\bx_i \}$ constitute transformations of the fixed
point through the $n$ intermediate mappings
$\phi_1(\cdot),\ldots,\phi_n(\cdot)$:
\begin{align}
   \tilde\bx_{1} & =\phi_1(\hat\bx)\\
   \tilde\bx_{2} & =\phi_2(\tilde\bx_{1})\\
   \cdots\\
   \tilde\bx_{n} & = \phi_n(\tilde\bx_{n-1}).  \\[-11mm]  
\end{align}
$ \hfill\blacksquare$ \\[1mm] The NC$^3$T shows that, in the limit
$t\to\infty$, the root-mean square value of ${\norm{\bx_t - \hat\bx}}$
is no larger than $\hat d$.

\subsubsection{Proof of the NC$^3$T}

First define the following shorthands:
\begin{align}
  \nu & = \sqrt{c^2 + \omega^2} , \\
  \sigma' & =  \omega M + \sigma .
\end{align}
Consider the following quadratic equation:
\begin{align}\label{eq:2ndordereq_nc3t}
  \hat d^2 =\nu^2  \hat d^2 +  2\sigma'\omega \hat d + \sigma'^2 .
\end{align}
Equation~\eqref{eq:2ndordereq_nc3t} has the unique non-negative
solution stated in \eqref{eq:dhatupbnd}:
\begin{align}
  \hat d & = \frac{\sigma'\omega}{ {1-\nu^2}} + \sqrt{ \frac{\sigma'^2\omega^2}{(1-\nu^2)^2} + \frac{\sigma'^2}{1-\nu^2}} \\
  & = \sigma'\cdot\frac{\omega + \sqrt{1-c^2}}{1-c^2-\omega^2} \\
  & = \sigma'\cdot\frac{\omega + \sqrt{1-c^2}}{ \pp{\sqrt{1-c^2}-\omega} \pp{ \sqrt{1-c^2}+\omega} } \\
  & = \frac{\sigma' }{\sqrt{1-c^2}-\omega} 
  = \frac{\omega M + \sigma  }{\sqrt{1-c^2}-\omega} .
\end{align}
 
Now define the error, 
\begin{align}
  d_{t,i} & = \sqrt{ \E{ \snorm{\bx_{t,i} - \tilde\bx_{i}}} } ,
\end{align}
and consider its evolution with $t$ and $i$.  Let   ${\tilde \bx_0 = \hat\bx}$.
For given $t$, we have for $i=0,\ldots,n-1$ that
\begin{align}
  d_{t,i+1}^2 & = \E{ \snorm{ \bx_{t,i+1} - \tilde\bx_{i+1}}} \\
  & = \E{ \snorm{ \phi_{i+1}(\bx_{t,i}) - \phi_{i+1}(\tilde\bx_{i}) + \beps_{t,i} } } \\
  & = \E{ \snorm{ \phi_{i+1}(\bx_{t,i}) - \phi_{i+1}(\tilde\bx_{i}) }} + \E{ \snorm{\beps_{t,i}} } \\
  & \qquad + 2\E{ (\phi_{i+1}(\bx_{t,i}) - \phi_{i+1}(\tilde\bx_{i}))^\T \beps_{t,i} } \\
  & \eqa \E{ \snorm{ \phi_{i+1}(\bx_{t,i}) - \phi_{i+1}(\tilde\bx_{i}) }} + \E{ \snorm{\beps_{t,i}} } \\
  & \leb c^2 \E{ \snorm{ \bx_{t,i} - \tilde\bx_{i} } } +  \E{ \snorm{\beps_{t,i}} }  \\
  & \lec c^2 \E{ \snorm{ \bx_{t,i} - \tilde\bx_{i} } } +  (\sigma' + \omega d_{t,i})^2 \\
  & = c^2 d_{t,i}^2 +  (\sigma' + \omega d_{t,i})^2  \\
  & =  \nu^2   d_{t,i}^2 +  2\sigma' \omega   d_{t,i} + \sigma'^2, \label{eq:dtdyn1}
\end{align}
where in (a) we used \eqref{eq:noise1stmom}, in (b) we used
\eqref{eq:contraction} and in (c) we used the following:
\begin{align}
  &  \E{\snorm{\beps_{t,i}}} =   \E{\E{ \snorm{\beps_{t,i}} | \bx_{t,i} }}  \\ 
  & \lea \E{ \pp{ \sigma + \omega \norm{\bx_{t,i}}}^2  } \\
  & = \E{\pp{ \sigma + \omega \norm{\bx_{t,i}-\tilde\bx_i + \tilde\bx_i}  }^2 } \\
  & \le \E{\pp{ \sigma + \omega \norm{\bx_{t,i}-\tilde\bx_i} + \omega\norm{\tilde\bx_i}  }^2 } \\
  & \le \E{\pp{ \sigma' + \omega \norm{\bx_{t,i}-\tilde\bx_i}  }^2 } \\
  & = \sigma'^2 + \omega^2 \E{\snorm{\bx_{t,i} - \tilde\bx_i }}
  +2\sigma'\omega \E{ {\norm{\bx_{t,i} - \tilde\bx_i}}} 
\end{align}
\begin{align}  
  &  \leb \sigma'^2 + \omega^2 \E{\snorm{\bx_{t,i} - \tilde\bx_i }}
  +2\sigma'\omega \sqrt{ \E{ \snorm{\bx_{t,i} - \tilde\bx_{i}}} } \\&
  =    (\sigma' + \omega d_{t,i})^2  . \label{eq:jensen1}
\end{align}
In \eqref{eq:jensen1}, (a) follows from \eqref{eq:noise2ndmom} and (b)
follows from Jensen's inequality and the concavity of $\sqrt{\cdot}$.

Consider \eqref{eq:dtdyn1}.
If $d_{t,i}\le \hat d$, then
\begin{align}
  d_{t,i+1} & \le  \sqrt{\nu^2 d_{t,i}^2 + 2\sigma' \omega d_{t,i} + \sigma'^2} \\
  & \le \sqrt{ \nu^2 {\hat d^2} + 2\sigma' \omega \hat d + \sigma'^2 } = \hat d.
\end{align}
If $d_{t,i} \ge \hat d$, then
\begin{align}
  d_{t,i+1} & \le \nu d_{t,i} + \pp{ \sqrt{\nu^2 d_{t,i}^2 + 2\sigma' \omega d_{t,i} + \sigma'^2} - \nu d_{t,i} } \\
  & \lea \nu d_{t,i} + \pp{ \sqrt{\nu^2 {\hat d^2} + 2\sigma' \omega \hat d + \sigma'^2} - \nu \hat d } \\
  & = \nu d_{t,i} + \hat d - \nu \hat d \\
  & = \hat d + \nu (d_{t,i} - \hat d),
\end{align}
where in (a) we used that ${ { \sqrt{\nu^2 d^2 + 2\sigma' \omega d +
      \sigma'^2} - \nu d } }$ is a decreasing function of $d$; this is
so because its derivative,
\begin{align}
     \frac{\nu^2 d+\sigma' \omega}{\sqrt{\nu^2 d^2   + 2\sigma' \omega d + \sigma'^2}} - \nu,
\end{align}
is non-positive if ${ (\nu^2 d + \sigma'\omega)^2 \le
  \nu^2\pp{\nu^2d^2 + 2\sigma' \omega d + \sigma'^2}}$, which always
holds.

Induction over $i$ gives that
\begin{align}
  d_{t,n}\le \hat d+ \nu^{n} (d_{t,0} - \hat d)^+.  \label{eq:dtn1}
\end{align}
Since $\hat\bx$ is the fixed point of $\varphi(\cdot)$, we have ${
  \tilde\bx_{n} = \hat\bx ,}$ which implies that
\begin{align}
  d_{t+1,0}=d_{t,n}.  \label{eq:dtn2}
\end{align}
Using \eqref{eq:dtn1} and \eqref{eq:dtn2} with induction over $t$,
\eqref{eq:nc3tmain} follows.  $\hfill\blacksquare$

\subsubsection{Remarks}

\begin{itemize}
\item In the special case of $n=1$, the NC$^3$T simplifies to a noisy
  version of the C$^2$T.  Suppressing superfluous subscripts, the
  recursion is
  \begin{align}
    \bx_{t+1} = \phi(\bx_t) + \beps_t,
  \end{align}
  $\tilde\bx_1=\hat\bx$ is the fixed point of $\phi(\cdot)$, and
  ${M=\norm{\hat\bx}}$.

  We refer to this special case as the noisy C$^2$T (NC$^2$T).

\item In the noise-free case, ${\sigma=\omega=0}$, so ${\hat d=0}$ and
  ${\sqrt{ \E{ \snorm{\bx_t - \hat\bx}} } =\norm{\bx_t - \hat\bx}}$.
  Then \eqref{eq:nc3tmain} reduces to \eqref{eq:FPTconvc3}, so the
  C$^3$T follows as a special case of the NC$^3$T.
  
\item If $\{\phi_i(\cdot)\}$ are contractions with \emph{different}
  factors $\{ c_i \}$, then the NC$^3$T holds with
  ${c=\max\{c_1,\ldots,c_n\}}$.

\item When $\omega>0$, the conditional noise variance scales with
  $\snorm{\bx_{t,i}}$.  This slows down the convergence, increasing the
  contraction factor from ${c}$ to ${\sqrt{c^2+\omega^2}}$.
  
  In fact, if $c^2+\omega^2\geq 1$, the mean-square error may diverge
  even if the noise-free recursion is a contraction.  For example,
  consider a scalar $x$, take $n=1$, and ${\phi_1(x)=x/2}$.  Suppose
  $\epsilon_t$ is selected such that ${\E{\epsilon_t | x}=0}$ and
  ${\E{\epsilon_t^2 | x} = (1+\sqrt{3}/2|x|)^2}$. Then with $\sigma=1$
  and ${\omega=\sqrt{3}/2}$,
  \eqref{eq:noise1stmom}--\eqref{eq:noise2ndmom} are satisfied with
  equality.  It follows that $c=1/2$, ${\hat x=0}$, and
  ${c^2+\omega^2=1}$, so equality holds in (b) in \eqref{eq:dtdyn1}
  and $d_{t,1}$ grows without bound:
\begin{align}
  d_{t+1,1}^2 & \ge c^2 d_{t+1,0}^2 + \sigma^2 + \omega^2 d_{t+1,0}^2 \\
  & = d_{t+1,0}^2+1  = d_{t,1}^2+1.
\end{align}
Hence, strict inequality in \eqref{eq:comegalt1} is crucial.
 
\item Suppose \eqref{eq:noise2ndmom} holds nominally and that the
  coordinate system is shifted by adding a constant vector $\ba$ to
  all $\{\bx_t\}$ and $\{\bx_{t,i}\}$.  Since
  \begin{align}
    \sigma + \omega \norm{\bx_{t,i}+\ba}
    \le \sigma + \omega \norm{\bx_{t,i}} + \omega\norm{\ba},
  \end{align}
  it follows that \eqref{eq:noise2ndmom} holds in the new coordinate
  system with $\omega$ unchanged, and $\sigma$ replaced with ${\sigma +
    \omega\norm{\ba}}$. Importantly, the contraction speed is
  unaffected.

\item Note that
  \begin{align}
    \sqrt{\sigma^2 + \omega^2 \snorm{\bx_{t,i}}} \le \sigma + \omega \norm{\bx_{t,i}}.
  \end{align} 
  Hence, the following alternative condition on
  the noise variance (used, for instance, in \cite{chen2012diffusion}),
  \begin{align}
    \E{ \snorm{\beps_t} | \bx_{t,i} } \le \sigma^2 + \omega^2 \snorm{\bx_{t,i}} \label{eq:altvarcond}
  \end{align}
   implies \eqref{eq:noise2ndmom} in the NC$^3$T and hence is more
   restrictive.
\end{itemize}

\subsection{Proving Convergence to a Desired Goal}

For iterative algorithms whose updates are contraction mappings, we
can use the C$^2$T, C$^3$T, and NC$^3$T to bound the distance between
$\bx_t$ and a desired goal $\bx_*$ (typically, the minimizer of a
global objective function).  In the noise-free case,
\begin{align}
   \norm{\bx_t - \bx_*} & = \norm{ \bx_t - \hat\bx + \hat \bx - \bx_*} \\
 & \lea  \norm{\bx_t-\hat\bx} + \norm{\hat\bx-\bx_*} \\
 & \leb c^{nt} \norm{\bx_0-\hat\bx} + \norm{\hat\bx-\bx_*}  , \label{finalbound2}
\end{align}
where (a) follows from the triangle inequality, and (b) follows from
the C$^2$T or C$^3$T.  In the noisy case,
\begin{align}
  & \qquad  \sqrt{ \E{ \snorm{\bx_t -\bx_*}} }  
   = \sqrt{ \E{   \snorm{\bx_t - \hat\bx + \hat\bx - \bx_*}} } \\ 
 &  \lea
  \sqrt{ \E{ \snorm{\bx_t- \hat\bx}} } 
  +\norm{ \hat\bx- \bx_*}  \\
  & \leb \hat d+\norm{ \hat\bx- \bx_*} + \sqrt{c^2 + \omega^2}^{\,nt} \pp{  {  \norm{\bx_0 - \hat\bx}} - \hat d}^+
  \\  
&
  \lec \hat d+\norm{ \hat\bx- \bx_*} \\ & \quad + \sqrt{c^2 + \omega^2}^{\,nt} \pp{   { \norm{\bx_0 - \bx_*} }
  +\norm{ \hat\bx- \bx_*}
  - \hat d}^+,
  \label{finalbound}
\end{align}
where (a) and (c) follow from the Minkowski inequality, and (b)
follows from the NC$^3$T.  

Equations~\eqref{finalbound2} and \eqref{finalbound} show the explicit
dependence of ${\norm{\bx_t-\bx_*}}$ on the initial error
${\norm{\bx_0 - \bx_*}}$, the linear convergence component, and the
error between the fixed point and the desired goal, ${\norm{ \hat\bx-
    \bx_*}}$.  This suggests a three-step approach, to be followed
when analyzing DGD and diffusion later in the paper: 1) establishing
and characterizing the contraction; 2) bounding the error between the
fixed point and the desired goal, ${\norm{ \hat\bx- \bx_*}}$; and 3)
using \eqref{finalbound2} or \eqref{finalbound}.
  
\subsection{Strongly Convex and Smooth Functions}

Throughout, we will be working with differentiable, strongly convex,
and smooth functions. Several equivalent characterizations of such
functions are available \cite[Ch.~2]{nesterov2018lectures}.

\paragraph*{First-order characterization}

A differentiable function ${f:\realR^m \smash{\to}\realR}$ is
$\mu$-strongly convex if
\begin{align}
  \mu \snorm{\by-\bx} \le  (\by-\bx)^\T (\nabla f(\by) - \naf)
  \label{eq:SCdef2}
\end{align}
for all $\bx$ and $\by$, and $L$-smooth if
\begin{align}
  \Vert\nabla f(\by) - \naf\Vert\leq L\Vert\by-\bx\Vert, \label{eq:SCdef1a}
\end{align}
for all $\bx$ and $\by$.

Note that the application of Cauchy-Schwarz inequality to
\eqref{eq:SCdef2} yields
\begin{align}
  \Vert\nabla f(\by) - \naf\Vert\geq \mu\Vert\by-\bx\Vert. \label{eq:SCdef1b}
\end{align}

\paragraph*{Second-order characterization}

A \emph{twice} differentiable function   ${f:\realR^m
  \to\realR}$ is $\mu$-strongly
convex and $L$-smooth if  
\begin{align}
  L \bI & \succcurlyeq \nabla^2 f(\bx) \succcurlyeq  \mu \bI   \label{eq:SCdef3}
\end{align}
for all $\bx$.  Equation \eqref{eq:SCdef3} states that $f(\cdot)$ lies
between two quadratic functions with curvatures $\mu$ and $L$,
respectively.

\subsection{The Mean Hessian Theorem\label{sec:mht}}

The gradients of $\mu$-strongly convex, $L$-smooth functions satisfy
the following important relation.\\[2mm]
\noindent\textbf{Mean Hessian Theorem (MHT):} Let ${f:\realR^m \to\realR}$ be
differentiable, $\mu$-strongly convex, and $L$-smooth.  Then for any
$\bx$, $\by$, there exists a symmetric matrix $ \bA_{\bx\by}$ (dependent on
$\bx$ and $\by$) such that
\begin{align}
  L \bI & \succcurlyeq \bA_{\bx\by} \succcurlyeq  \mu \bI , \label{eq:ldpr1} \\
 \nabla f(\by) - \naf & = \bA_{\bx\by}(\by-\bx).     \label{eq:ldpr2} 
\end{align}
$ \hfill\blacksquare$

We will prove the MHT in two different ways: 
\begin{enumerate}
\item[(i)] first, using a line integral argument assuming twice
  differentiability of $f(\cdot)$; then
\item[(ii)] again, using linear algebra assuming \emph{only (once)}
  differentiability.
\end{enumerate}
Case (ii) of course subsumes case (i), rendering the proof for case
(i) redundant -- however, this proof is so intuitive that we give it
independently.

Note that in case (ii) the Hessian may not exist, so strictly speaking
MHT is a misnomer.  Also note that $\bA_{\bx\by}$ may not be unique.
For example, \eqref{eq:ldpr2} remains satisfied if incrementing or
decrementing $\bA_{\bx\by}$ by the Gramian of any vector that is
orthogonal to $\by-\bx$.

\subsubsection{Proof of MHT for Twice Differentiable  $f(\cdot)$}\label{sec:twicediff}

If $f(\cdot)$ is twice differentiable everywhere, the MHT can be
established by integrating $\nabla^2 f(\cdot)$ along a line.  Let
\begin{align}
  \bg(\xi) = \nabla f((1-\xi)\bx + \xi\by).
\end{align}
Then
\begin{align}
  \bg(0) & =  \naf, \\
  \bg(1) & =  \nabla f(\by), \\
  \bg'(\xi) & = [\nabla^2 f((1-\xi)\bx + \xi\by)](\by-\bx),
\end{align}
wherefrom
\begin{align}
  \nabla f(\by) & = \bg(1)  =
  \bg(0) + \int_0^1 d\xi \; \bg'(\xi) \\
  & =   \naf +\left[ \int_0^1 d\xi \; \nabla^2 f((1-\xi)\bx + \xi\by) \right] (\by-\bx). \label{eq:MHT}
\end{align}
Hence \eqref{eq:ldpr2} is satisfied with
\begin{align}\label{eq:meanhess}
 \bA_{\bx\by}= \int_0^1 d\xi \; \nabla^2 f((1-\xi)\bx + \xi\by),
\end{align}
which equals the mean Hessian between $\bx$ and $\by$. 
Owing to \eqref{eq:SCdef3},
\begin{align}\label{eq:mvt2u1}
  L \ge \frac{\bz^\T \ppb{ \nabla^2 f((1-\xi)\bx + \xi \by)} \bz}{\snorm{\bz}} \ge  \mu  ,
\end{align}
for all $\xi$ and all  $\bz$, $\bz\neq\bzero$.  Integration of
\eqref{eq:mvt2u1} with respect to $\xi$ over $[0,1]$ gives
\begin{align}
  L \ge \frac{\bz^\T  \bA_{\bx\by}  \bz}{\snorm{\bz}} \ge  \mu  ,
\end{align}
for all  $\bz\neq\bzero$, so \eqref{eq:ldpr1} is satisfied.

\subsubsection{Proof of MHT for (Only Once) Differentiable  $f(\bx)$}\label{sec:oncediff}
 
To show the MHT when $f(\cdot)$ is (only) once differentiable
everywhere, let
\begin{align}
  \ba & = \by-\bx,\\
  \bb & = \nabla f(\by) - \naf. 
\end{align}
If $\bx=\by$ there is nothing to prove, so assume $\bx\neq \by$; then
$\ba\neq\bzero$ and from \eqref{eq:SCdef1b}, $\bb\neq\bzero$.  From
\eqref{eq:SCdef1a} and \eqref{eq:SCdef1b} we have
\begin{align}
  \mu\norm{\ba}\le\norm{\bb} \le L \norm{\ba}. \label{eq:mhtproof1a}
\end{align}
Hence,
\begin{align}
  \mu\bb^\T\ba \lea \mu\norm{\ba}\norm{\bb} \le \snorm{\bb}, \label{eq:mhtproof1b}
\end{align}
where (a) follows from the Cauchy-Schwarz inequality.

Equality in (a) in \eqref{eq:mhtproof1b} is only possible if $\ba$ and
$\bb$ are parallel; ${\ba=\norm{\ba}/\norm{\bb}\cdot\bb}$.  In this
case, take ${\bA_{\bx\by} = \norm{\bb}/\norm{\ba}\cdot\bI}$. Then,
because of \eqref{eq:mhtproof1a}, \eqref{eq:ldpr1} is satisfied. Since
\begin{align}
  \bA_{\bx\by}\ba
  = \bA_{\bx\by}\frac{\norm{\ba}}{\norm{\bb}}\bb
  = \frac{\norm{\bb}}{\norm{\ba}}\frac{\norm{\ba}}{\norm{\bb}}\bb
  = \bb,
\end{align}
\eqref{eq:ldpr2} is satisfied too.

Now suppose $\ba$ and $\bb$ are \emph{not} parallel, so that 
strict inequality holds in (a) in \eqref{eq:mhtproof1b}.
Then ${ \snorm{\bb} -  \mu\bb^\T\ba >0  }$ and we can define
\begin{align}
  \alpha & =  \frac{ \bb^\T \ba  -  \mu \snorm{\ba}}{\snorm{\bb}  - \mu \bb^\T\ba} . \label{eq:defalphamht}
 \end{align}
The use of Theorem~2.1.12 in \cite{nesterov2018lectures} gives that
${\bb^\T \ba - \mu \snorm{\ba} \ge (\snorm{\bb} - \mu \bb^\T\ba)/L}$,
which implies ${\alpha\ge 1/L}$.  Furthermore, since $\ba$ and $\bb$
are not parallel we must have ${\norm{\bb-\mu\ba}> 0}$, which implies
${ \snorm{\bb}+\mu^2\snorm{\ba}-2\mu\bb^\T\ba>0}$, and in turn that
$\alpha<1/\mu$.  Take
\begin{align}
  \bA_{\bx\by} & = \mu\Pi_{(\bb  - \mu \ba)}^\perp  + \frac{1}{\alpha} \Pi_{(\bb  - \mu \ba)}.  \label{eq:ldpr3}
\end{align}
Clearly, all eigenvalues of $\bA_{\bx\by} $ lie between $\mu$ and $L$,
so \eqref{eq:ldpr1} is satisfied.  It remains to show that
\eqref{eq:ldpr2} holds:
\begin{align}
  \bA_{\bx\by} \ba & = \mu\Pi_{(\bb  - \mu \ba)}^\perp\ba  + \frac{1}{\alpha} \Pi_{(\bb  - \mu \ba)}\ba \\
  & = \mu\ba + \pp{\frac{1}{\alpha} - \mu} \Pi_{(\bb  - \mu \ba)} \ba \\
  & = \mu\ba + \pp{\frac{1}{\alpha} - \mu} \frac{({\bb  - \mu \ba})^\T\ba}{\snorm{{\bb  - \mu \ba}}} (\bb-\mu\ba) \\
  & = \mu\ba + \pp{\frac{1}{\alpha} - \mu} \frac{\bb^\T\ba-\mu \snorm{\ba}}{\snorm{\bb-\mu\ba}} (\bb-\mu\ba) \\
  & = \mu\ba + \pp{\frac{1}{\alpha} - \mu} \frac{\alpha}{1-\alpha\mu} (\bb-\mu\ba) \\
  & = \mu\ba + \bb-\mu\ba = \bb. \\[-11mm]
\end{align}
$ \hfill\blacksquare$

\subsubsection{Misconceptions}

Consider the twice-differentiable case.  Equation \eqref{eq:MHT} is
reminiscent of the mean value theorem for scalar, differentiable
functions
\cite[Th.~5.10]{rudin1953principles}\footnote{\setstretch{0.7}If $x\le
z \le y$, then ${g(y)=g(x) + g'(z)(y-x)}$.}; in fact, if $\bx$ is
scalar then $\bA_{\bx\by}$ is scalar too and \eqref{eq:ldpr2} is
simply this mean value theorem applied to $f'(\cdot)$.  Given this, one
may intuit that there exists a $\xi$ such that ${ \nabla f(\by) =
  \nabla f(\bx) + [\nabla^2 f( (1-\xi) \bx + \xi \by )] (\by-\bx) }$.
But this is \emph{not} the case, and the assumed existence of such a
$\xi$ appears to be a common misconception. For example, this fallacy
can be found in papers with analyses of GD algorithms
\cite[Lemma~5]{yang2020energy} and, as pointed out in
\cite{feng2013mean}, in statistics books and papers.

\section{Gradient Descent (GD)}\label{sec:GD}

As a preparation for the analysis of DGD and diffusion, we first
review the standard GD algorithm for the minimization of a
differentiable, $\mu$-strongly convex, and $L$-smooth function
${f:\realR^m\to \realR}$, and show how the MHT+contraction framework
yields a tight convergence bound. 

GD minimizes $f(\cdot)$ by computing the sequence of iterates
$\{\bx_t\}$ according to
\begin{align}\label{eq:GDiter}
  \bx_{t+1} =   \bx_t - \eta \nabla f(\bx_t) 
\end{align}
for some initial point $\bx_0$, and positive step size $\eta$.  Let
\begin{align}
  \bx_*=\argmin_\bx f(\bx)
\end{align} 
be the optimal point, which is unique because of the strong convexity,
and at which $ \nafs=\bzero $.

To investigate under what conditions $\bx_t\to\bx_*$ as $t\to\infty$,
define the mapping
\begin{align}
 \phi(\bx) = \bx - \eta\naf,\label{eq:defphiGD}
\end{align}
so that $\bx_{t+1}=\phi(\bx_t)$. Then for any vectors $\bx$ and $\by$,
\begin{align}
  & \quad  \norm{ \phi(\by) - \phi(\bx) } \\
  &  = \norm{ \by -  \eta\nabla f(\by) - [\bx -  \eta\naf]} \\
  & \eqa \norm{ \by - \eta[\naf + \bA_{\bx\by}(\by-\bx)] -   [\bx -  \eta\naf]} \\
  & = \norm{ (\bI - \eta\bA_{\bx\by})(\by-\bx) } \\
  & \le \mnorm{\bI - \eta\bA_{\bx\by}}  \norm{ \by - \bx } \\
  & \le \max\{ |1-\eta \mu|, |1-\eta L| \} \cdot\norm{ \by - \bx } \\
  & \eqb
  \begin{cases}
    (1-\eta \mu)   \norm{ \by - \bx } ,   & 0<\eta\le \frac{2}{L+\mu}, \\
    (\eta L - 1)   \norm{ \by - \bx } ,   & \frac{2}{L+\mu}\le \eta  .
  \end{cases}     \label{eq:GDcontrnew2}
\end{align}
In step (a), we applied the MHT to $f(\cdot)$: for any $\bx$ and $\by$
there is a matrix $\bA_{\bx\by}$ that satisfies \eqref{eq:ldpr1} and
\eqref{eq:ldpr2}.  To establish (b), note that ${\max\{ |1-\eta \mu|,
  |1-\eta L| \} = 1-\eta\mu}$ if and only if ${ (1-\eta\mu)^2 \le
  (1-\eta L)^2 }$, or equivalently,
\begin{align}\label{eq:etacondition}
  \eta \le \frac{2}{L+\mu}.
\end{align}
Hence, if ${0 < \eta\le 2/(L+\mu)} $, $\phi(\cdot)$ is a contraction
with factor ${1-\eta\mu<1}$, and if $ {2/(L+\mu) \le \eta < 2/L }$,
$\phi(\cdot)$ is a contraction with factor ${\eta L-1<1}$.  The
smallest possible contraction factor equals ${ (L-\mu)/(L+\mu) }$ and
is obtained with ${\eta=2/(L+\mu)}$. Note that any contraction factor
between ${ (L-\mu)/(L+\mu) }$ and $1$ can be obtained for two
different values of $\eta$: one below ${2/(L+\mu)}$ and one above
${2/(L+\mu)}$.

Let $\hat\bx$ be the fixed point of $\phi(\cdot)$.  By inspection of
\eqref{eq:defphiGD}, $\phi(\bx_*)=\bx_*$, so the fixed point equals
the optimal point: ${\hat\bx=\bx_*}$.  By using the C$^2$T, we
conclude that
\begin{align}
  \hspace*{-5mm} \norm{ \bx_t - \bx_* }  &  \le 
  \begin{cases}
    (1-\eta \mu)^t   \norm{ \bx_0 - \bx_* }, & \eta\leq \frac{2}{L+\mu}, \\
    (\eta L-1)^t   \norm{ \bx_0 - \bx_* }, & \frac{2}{L+\mu}\le \eta<\frac{2}{L}.
  \end{cases}\label{eq:GDcontrnewe}
\end{align}
Hence $\bx_t \to \bx_*$ as $t\to\infty$, geometrically fast -- that
is, GD achieves linear convergence if $\eta<2/L$.

\paragraph*{Tightness of the Bound \eqref{eq:GDcontrnewe}}

To illustrate the tightness of the bound \eqref{eq:GDcontrnewe},
consider GD for $m=2$, with predetermined $\eta$ and $\bx_0$, applied
to ${f(\bx) = \mu x_1^2/2 + L x_2^2/2}$.  By direct calculation,
${\bx_*=\bzero}$, and
\begin{align}
  \bx_t & = \begin{bmatrix} 1-\eta\mu & 0 \\ 0 & 1 - \eta L \end{bmatrix}^t \bx_0.  \label{eq:GDexample1}
\end{align}
If ${ \eta \ge 2/L }$, then with ${\bx_0=[0,1]^\T}$, ${\bx_t}$
oscillates between ${[0,1]^\T}$ and ${[0,-1]^\T}$, or diverges.  If ${
  2/(L+\mu)\le \eta <2/L }$, the worst-case choice of $\bx_0$ is ${
  \bx_0=[0,1]^\T }$, which gives ${ \norm{\bx_t-\bx_*} = (\eta L-1)^t
  \norm{\bx_0-\bx_*} }$.  Finally, if ${\eta\le 2/(L+\mu)}$, the
worst-case $\bx_0$ is ${ \bx_0=[1,0]^\T }$, and ${ \norm{\bx_t-\bx_*}
  = (1-\eta \mu)^t \norm{\bx_0-\bx_*} }$.  Taken together, this shows
that for a given step size ${\eta<2/L}$, there exist a worst-case
function $f(\cdot)$ and initialization $\bx_0$ such that the bound in
\eqref{eq:GDcontrnewe} is met with equality.  {Conversely,
  when ${\eta\geq 2/L}$, GD either does not converge or diverges, in
  the worst case.  We conclude that \eqref{eq:GDcontrnewe}
  characterizes the GD contraction over the entire range of $\eta$ for
  which convergence is guaranteed.}

We shall be mainly interested in operating regimes where $\eta$ can be
arbitrarily small.  For GD, the relevant step size condition is then
\eqref{eq:etacondition}, under which \eqref{eq:GDcontrnewe} gives
\begin{align}
  \norm{ \bx_t - \bx_* }   \le (1-\eta \mu)^t   \norm{ \bx_0 - \bx_* } .  \label{eq:GDcontrnewx}
\end{align}
Note that the contraction factor in \eqref{eq:GDcontrnewx},
${1-\eta\mu}$, is only dependent on $\mu$ but not on $L$; however, the
condition on $\eta$ in \eqref{eq:etacondition} depends on $L$.  To the
authors' knowledge, existing derivations of the bound in
\eqref{eq:GDcontrnewx} with contraction factor ${1-\eta\mu}$ rely on
\emph{first-order characterizations} of strong convexity and
smoothness; see, for example \cite[Th.~3.12]{bubeck2015convex}.  In
contrast, our MHT+contraction approach uses a description akin to
\emph{second-order characterizations} of twice differentiable
functions, but without requiring twice differentiability.  The
powerfulness of this approach is that when bounding the eigenvalues of
${\bI-\eta\bA_{\bx\by}}$, the strong convexity and smoothness
properties of $f(\cdot)$ are used simultaneously, which immediately
gives \eqref{eq:GDcontrnew2}. Some analyses in the literature that use
these properties separately yield looser bounds; see, for example,
\cite[Th.~2 (Ch.~1)]{polyakbook},
\cite[Th.~3.10]{polyak2021introduction}, \cite[Lemma~1]{8786146}, and
\cite[Th.~1]{9515709}.

\paragraph*{Unboundedness of $\naf$}

A common misconception is that the gradients of strongly convex
functions can be uniformly bounded over $\realR^m$.  For example, this
assumption is made in \cite[Th.~14.11]{Shalev-Shwartz_Ben-David_2014},
and in
\cite{9515709,xing2021federated,10338845,liconvergence,stich2019,pmlr-v130-ruan21a,jadbabaie2023federated}
(for distributed GD applied to federated learning).  But as pointed
out for example in \cite{nguyen2019new}, a strongly convex function
always has unbounded gradients over $\realR^m$.  This is a simple
consequence of \eqref{eq:SCdef1b}: if $f(\cdot)$ is $\mu$-strongly
convex then ${\norm{\naf} =\norm{\naf - \nafs} \ge
  \mu\norm{\bx-\bx_*}}$, which is unbounded on $\realR^m$.

Importantly, our MHT+contraction analysis framework requires no
assumptions on boundedness of the gradients. 

\section{Decentralized Gradient Descent (DGD) and Diffusion}\label{sec:DGD}

This section contains the main analysis of DGD and diffusion. We first
describe the algorithms, and then analyze their convergence properties
using the MHT, C$^2$T, and C$^3$T, both for single and multiple local
gradient updates per iteration.

\subsection{Communication Network Topology}\label{subsec:topology}

We consider $N$ agents interconnected in a communication network
defined by its adjacency (weight) matrix $\bW$.  We assume that this
network is connected, undirected (${\bW^\T=\bW}$), and that the weights
sum up to one at each agent so that ${\bW\bone=\bone}$. We also assume
that the (real-valued) eigenvalues of $\bW$, $\{\lambda_n\}$, when
ordered in descending order, satisfy
\begin{align}
  1=\lambda_1 > \lambda_2 \ge \cdots \ge \lambda_N > -1.   \label{eq:Weigs}
\end{align}
Note that $\bW$ is \emph{not} required to be non-negative, and not
required to be positive (semi-)definite.  The case of ${\lambda_N=-1}$
is discussed in
Sections~\ref{sec:discuss}.\ref{sec:suffcond}--\ref{sec:requirementsW}.

Under the assumptions made, we have that $\bW^t\to\bone\bone^\T/N$ as
$t\to\infty$, which implies ${\bW^t\bx\to\bone\bone^\T\bx/N}$ for any
$\bx$. For this reason, the multiplication of a vector $\bx$ with
$\bW$ is known as \emph{consensus averaging}.

\subsection{The DGD and Diffusion Algorithms}

All relevant structural aspects of the problem surface for a scalar
parameter, considered here.  Let ${f_n:\realR \to \realR}$ be the
local objective function of the $n$th agent; herein, we simply treat
each $f_n(\cdot)$ as a given function while in an application, it
would be dependent on data available at the $n$th agent.  We assume
that all $\{f_n(\cdot)\}$ are $\mu$-strongly convex and $L$-smooth.
The goal is to minimize the global objective function
\begin{align}
  F(x) = \frac{1}{N} \sum_{n=1}^N f_n(x), \label{eq:globalF}
\end{align}
which   has the unique minimum
\begin{align}
  x_* = \argmin_x F(x),
\end{align}
at which $F'(x_*)=0$.

For future use, we introduce the following ``component-wise''
objective function
\begin{align}
  f(\bx) = \sum_{n=1}^N f_n(x_n),  \label{eq:deffbx}
\end{align}
where ${\bx = [x_1,...,x_N]^\T}$, and define
\begin{align}
  \bx_* = x_*\bone = [x_*, \ldots, x_* ]^\T.
\end{align}
(This vector $\bx_*$ must not be conflated with $\bx_*$ in
Section~\ref{sec:GD}; henceforth we write $\bx_*$ in lieu of
$x_*\bone$ for brevity.)  Note that
\begin{align}
  \bW\bx_* & = \bx_*, \label{eq:xstarnull}\\
  \naf & = [    f'_1(x_1) ,..., f'_N(x_N)]^\T , \\
  \bone^\T \nafs & = \sum_{n=1}^N f'_n(x_*) = N F'(x_*) = 0, \label{eq:bonenafs}
\end{align}
but that $ \nafs \neq \bzero$ in general.

DGD and diffusion work by having each agent $n$ forming a sequence of
iterates. We let $\bx_t$ be the $N$-vector whose $n$th component
represents the iterate of the $n$th agent at time $t$.  The iterates
are formed by consensus averaging among the agents' neighbors using
the weights in $\bW$, and subtraction of a local gradient update.
There are three basic versions of the algorithms:
\begin{itemize}
\item \emph{Standard DGD} (called ``consensus strategy'' in
  \cite{6777576}):
  \begin{align}
    \bx_{t+1} & =  \bW\bx_t -  \eta \nabla f(\bx_{t}) , \label{eq:stdDGD}  
  \end{align}
where $\eta$ is a positive step size.  Here each agent computes its
local gradient update based on the iterate available at the beginning
of the iteration.  This allows agents to compute their gradient
updates while simultaneously communicating with their neighbors.
  
\item \emph{Adapt-then-combine (ATC) diffusion} (e.g.,
  \cite{8491372,SAYED2014323}):
  \begin{align}
    \bx_{t+1} = \bW (\bx_t - \eta \nabla f(\bx_t)).\label{eq:diffusionatc}
  \end{align}
  Here too, each agent computes its local gradient update based on the
  iterate available at the beginning of the iteration, but in contrast to
  the standard DGD, these updates are weighted by $\bW$.

\item \emph{Combine-then-adapt (CTA) diffusion} (e.g., \cite{SAYED2014323}), 
  \begin{align}
    \bx_{t+1} = \bW \bx_t - \eta \nabla f(\bW\bx_t).\label{eq:diffusioncta}
  \end{align}
  Here, each agent computes its gradient update \emph{after} forming
  consensus with the incoming iterates from the neighbors.
\end{itemize}

ATC and CTA diffusion are closely interrelated. In fact, CTA diffusion
becomes equivalent to ATC diffusion if $\bx_t$ is re-defined to be the
output after applying $\bW$.  Therefore, we will henceforth only
consider the standard DGD in \eqref{eq:stdDGD}, and ATC diffusion in
\eqref{eq:diffusionatc}, and use the names \emph{DGD} and
\emph{diffusion} for these two algorithms.
Defining the matrix
\begin{align}
  \bZ & = \begin{cases}
    \bI, & \mbox{for DGD} \\
    \bW, & \mbox{for diffusion},
  \end{cases}  \label{eq:defZ}
\end{align}
we can write the DGD and diffusion iterations \eqref{eq:stdDGD} and
\eqref{eq:diffusionatc} as
\begin{align}
  \bx_{t+1} = \bW \bx_t - \eta \bZ \nabla f(\bx_t).  \label{eq:unifiter}
\end{align}

\subsection{Convergence to a Fixed Point}\label{sec:convfix}

We first show that for both DGD and diffusion the iterate $\bx_t$
converges geometrically to a fixed point (to be denoted
$\hat\bx$). Towards this end, define the mapping
\begin{align}
  \phi(\bx) = \bW\bx - \eta\bZ\naf,  \label{eq:defphidgd}
\end{align}
so that \eqref{eq:unifiter} can be written
\begin{align}
\bx_{t+1} & = \phi(\bx_t) .  \label{eq:DGDfix}
\end{align}

\subsubsection{[Standard] DGD}\label{sec:convstdDGD}

Consider DGD ($\bZ=\bI$). For any vectors $\bx$ and $\by$, 
\begin{align}
  & \qquad \norm{ \phi(\by) - \phi(\bx) } \\
  & = \norm{ \bW \by - \eta  \nabla f(\by) -  [\bW \bx - \eta  \nabla f(\bx)]  } \\
  & \eqa \norm{ \bW \by - \eta  [\nabla f(\bx) + \bA_{\bx\by}(\by - \bx)] - [\bW\bx - \eta\nabla f(\bx)] } \\
  & = \norm{ (\bW - \eta \bA_{\bx\by}) (\by - \bx ) } \\   
  & \leb  \max\{ |\lambda_N - \eta L|, |1-\eta \mu| \} \cdot\norm{ \by - \bx }  \\
  & =  \begin{cases}
    (1-\eta \mu)   \norm{ \by - \bx } ,   & 0<\eta\le \frac{1+\lambda_N}{L+\mu}, \\
    (\eta L - \lambda_N )  \norm{ \by - \bx } ,   & \frac{1+\lambda_N}{L +\mu}\le \eta  ,
  \end{cases}    \label{eq:DGDfix2}
\end{align}
where in (a) we applied the MHT to $f(\cdot)$ and in (b) we used 
\begin{align}  
  ( \lambda_N - \eta L )\bI   & \preccurlyeq
  \bW-\eta\bA_{\bx\by}        \preccurlyeq (1-\eta \mu) \bI  .
\end{align}
Hence $\phi(\cdot)$ is a contraction with factor $1-\eta\mu<1$ if
\begin{align}\label{eq:eta2}
  \eta \le \frac{1+\lambda_N}{L+\mu},
\end{align}
and a contraction with factor ${\eta L - \lambda_N <1}$ if ${
  (1+\lambda_N)/(L+\mu) \le \eta < (1+\lambda_N)/L }$.

\subsubsection{Diffusion}

For diffusion, $\bZ=\bW$. Then  for any $\bx$ and $\by$,
\begin{align}
  \norm{ \phi(\by) - \phi(\bx) }  & = \norm{ \bW \pp{ \by - \eta   \nabla f(\by) -  [  \bx - \eta   \nabla f(\bx)]  } }\\
  & \le \mnorm{ \bW} \norm{   \by - \eta   \nabla f(\by) -  [  \bx - \eta   \nabla f(\bx)]   }\\
  & {\eqa}   \norm{   (\bI - \eta \bA_{\bx\by}) (\by - \bx ) } \\
  & \leb    \begin{cases}
    (1-\eta \mu)   \norm{ \by - \bx } ,   & 0<\eta\le \frac{2}{L+\mu}, \\
    (\eta L - 1)   \norm{ \by - \bx } ,   & \frac{2}{L+\mu}\le \eta  ,
  \end{cases} \\
  \label{eq:diffusioncontractive1}
\end{align}
where in (a) we used that ${\mnorm{\bW}= 1}$ and in (b) we used the
calculation in Section~\ref{sec:GD}.  The conditions on $\eta$ for
$\phi(\cdot)$ to be a contraction are the same as for GD.

\subsubsection{Convergence to a Fixed Point}

For both DGD and diffusion, the C$^2$T shows that under the stated
conditions on $\eta$, the iteration \eqref{eq:unifiter} converges to
fixed point, $\hat\bx$ such that
\begin{align}
  \hat\bx = \bW \hat\bx- \eta \bZ \nabla f(\hat\bx) . \label{eq:fixedpunif1}
\end{align}
Specifically, for DGD:
\begin{align}
  \norm{\bx_t - \hat\bx} \le  \begin{cases}
    (1-\eta \mu)^t   \norm{ \bx_0 - \hat\bx } ,   & 0<\eta\le \frac{1+\lambda_N}{L+\mu}, \\
    (\eta L - \lambda_N )^t     \norm{ \bx_0 - \hat\bx } ,   & \frac{1+\lambda_N}{L +\mu}\le \eta < \frac{1+\lambda_N}{L} ,
  \end{cases} \label{eq:DGDboundnew}
\end{align}
and for diffusion: 
\begin{align}
  \hspace*{-5mm} \norm{\bx_t - \hat\bx} \le  \begin{cases}
    (1-\eta \mu)^t   \norm{ \bx_0 - \hat\bx } ,   & 0<\eta\le \frac{2}{L+\mu}, \\
    (\eta L -1 )^t     \norm{ \bx_0 - \hat\bx } ,   & \frac{2}{L +\mu}\le \eta < \frac{2}{L} .
  \end{cases} \label{eq:diffboundnew}
\end{align}

\subsubsection{Tightness of the Bounds \eqref{eq:DGDboundnew} and \eqref{eq:diffboundnew}}

To appreciate the tightness of these bounds, consider DGD applied to
${f_n(x)=\rho x^2 /2}$, for some $\rho>0$. Then ${\hat\bx=\bzero}$ and
\begin{align}
  \bx_t-\hat\bx = (\bW-\eta\rho\bI)^t (\bx_0-\hat\bx).
\end{align}
Let $\bu_n$ be a unit-norm eigenvector of $\bW$ associated with the
eigenvalue $\lambda_n$, and note that with the initial point
${\bx_0=\bu_n}$,
\begin{align}
  \bx_t-\hat\bx =  (\bW-\eta\rho\bI)^t \bu_n = (\lambda_n - \eta\rho)^t \bu_n.
\end{align}
If ${\eta\ge (1+\lambda_N)/L}$, ${\rho=L}$, and ${\bx_0=\bu_N}$,
$\bx_t$ either oscillates between $ \bu_N$ and $-\bu_N$, or diverges.
If ${(1+\lambda_N)/(L+\mu) \le \eta < (1+\lambda_N)/L}$, ${\rho=L}$,
and ${\bx_0=\bu_N}$, we have ${ \norm{\bx_t-\hat\bx} = (\eta L -
  \lambda_N)^t}$.  Finally, if ${ \eta \le (1+\lambda_N)/(L+\mu) }$,
${\rho=\mu}$, and ${\bx_0=\bu_1=\bone/\sqrt{N}}$, we have ${
  \norm{\bx_t-\hat\bx} = ( 1 - \eta\mu)^t}$.  This shows that for
arbitrary $\bW$, and a given step size $\eta<(1+\lambda_N)/L$, there
is a set of worst-case functions $\{f_n(\cdot)\}$ and an initial point
$\bx_0$ for which \eqref{eq:DGDboundnew} is met with equality.
{Conversely, when ${\eta\geq(1+\lambda_N)/L}$, DGD may either
  not converge, or diverge.}

A similar example can be constructed to demonstrate the tightness of
the bound for diffusion, \eqref{eq:diffboundnew}.

{We conclude that \eqref{eq:DGDboundnew} and
  \eqref{eq:diffboundnew} characterize the DGD and diffusion
  contractions over the entire range of $\eta$ for which convergence
  to a fixed point is guaranteed.}

\subsection{Distance from the Fixed Point to the Optimum} 

We next upper-bound the distance from the fixed point to the optimum,
${\norm{\hat\bx - \bx_*}}$. To this end, note from
\eqref{eq:fixedpunif1} that
\begin{align}
  \bzero   & =  \frac{1}{\eta} (\bI-\bW)\hat\bx + \bZ \nabla f(\hat\bx)  \\
  & \eqa  \frac{1}{\eta} (\bI-\bW)\hat\bx +  \bZ\nafs + \bZ\bA_{\bx_*\hat\bx} (   \hat\bx -  \bx_* ) \\
  & \eqb  \frac{1}{\eta} (\bI-\bW)(\hat\bx - \bx_*) +  \bZ\nafs + \bZ\bA_{\bx_*\hat\bx} (   \hat\bx -  \bx_* ) \\
  & =  \ppb{\bZ\bA_{\bx_*\hat\bx} + \frac{1}{\eta} (\bI-\bW)}(\hat\bx - \bx_*)  +  \bZ\nafs  ,   \label{eq:fixedpunif3}
\end{align}
where $\bA_{\bx_*\hat\bx}$ satisfies \eqref{eq:ldpr1}; in (a) we
applied the MHT to $f(\cdot)$, and in (b) we used
\eqref{eq:xstarnull}.  Now apply the distance-bounding result in
Appendix~\ref{subsubsec:auxbnd} to \eqref{eq:fixedpunif3}, with
${\bd=\hat\bx-\bx_*}$ and ${\bv=\nafs}$, and use
\eqref{eq:bonenafs}. This yields
\begin{align}
  \norm{ \hat\bx - \bx_*}    & \le    \eta  \frac{L}{\mu}\Lambda  \norm{\nafs}   , \label{eq:normxGxs} 
\end{align}
where we define the ``topology factor''
\begin{align}
  \Lambda & = \begin{dcases} \mnorm{\bIWp} = \frac{1}{1-\lambda_2}, & \mbox{for DGD} \\
    2 \mnorm{\bIWp\bW}, & \mbox{for diffusion},
  \end{dcases}  \label{eq:defLambda}
\end{align}
and where for diffusion we additionally require ${\eta\le 1 /(L \Lambda )}$.

The right hand side of \eqref{eq:normxGxs} consists of four factors:
$\eta$, which must be small for $\hat\bx$ to be close to $\bx_*$; the
condition number ${L/\mu}$, which depends on the convexity and
smoothness properties of the local objectives; $\Lambda$, which
differs between DGD and diffusion; and $\norm{\nafs}$ that captures
the data heterogeneity.

\subsection{Bounding the Total Error}\label{subsec:bxtbs}

Henceforth, we will be interested in criteria that guarantee
convergence for arbitrarily small $\eta$ and hence impose
\eqref{eq:eta2} for DGD and \eqref{eq:etacondition} for diffusion,
leading to
\begin{align}
  \norm{\bx_t - \hat\bx} \le  (1-\eta\mu)^t \norm{\bx_0 - \hat\bx}  \label{eq:fixedpunif2}
\end{align}
for both algorithms. {Readers interested in the ``upper step
  size regimes'' [${ (1+\lambda_N)/L>\eta\ge (1+\lambda_N)/(L+\mu)}$
    for DGD, and ${2/L>\eta\ge 2/(L+\mu)}$ for diffusion,
    respectively] may simply replace ${1-\eta\mu}$ by ${\eta L -
    \lambda_N } $ or ${\eta L-1}$ wherever pertinent in the subsequent
  development.}

From \eqref{finalbound2} and \eqref{eq:fixedpunif2}, we obtain a bound
on the total error:
\begin{align}
  \norm{\bx_t - \bx_*} & \le (1-\eta\mu)^t \norm{\bx_0 - \hat\bx} +   \norm{\hat\bx - \bx_*}. \label{eq:trieq}
\end{align}
For any given $\epsilon>0$, there exists $\eta$ such that the second
term of \eqref{eq:trieq} is less than $\epsilon /2$.  For that $\eta$,
there is a value of $t$ such that the first term is less than
$\epsilon /2$; thus ${\norm{\bx_t - \bx_*}\le \epsilon}$.  In
particular, $\eta$ should be of order $\sim\epsilon$ to make the
second term in \eqref{eq:trieq} small. Since ${1-\eta\mu\le
  e^{-\eta\mu}}$, $t$ should be of order ${\sim
  \log(1/\epsilon)/\epsilon }$ to make the geometric factor
${(1-\eta\mu)^t}$ small.

\subsection{Comparison Between DGD and Diffusion}

The convergence properties of DGD and diffusion have a complicated
dependence on the curvature of $\{f_n(\cdot)\}$ and on the network
topology.  Some intuition for the dependence on topology can be gained
by considering the distance $\norm{\bxhs}$ in \eqref{eq:normxGxs}.
This distance scales with $1/(1-\lambda_2)$ for DGD and with
$\mnorm{(\bI-\bW)^\dagger\bW}$ for diffusion.  Since
\begin{align}
  \mnorm{(\bI-\bW)^\dagger\bW} \leq \mnorm{(\bI-\bW)^\dagger} = \frac{1}{1-\lambda_2},  \label{eq:dgddiffcomp}
\end{align}
diffusion has a more favorable dependence on the topology than DGD.

While DGD operates under the step size condition \eqref{eq:eta2},
diffusion requires the condition
\eqref{eq:etacondition}. Additionally, for diffusion,
\eqref{eq:normxGxs} requires ${\eta\le 1 /(L \Lambda )}$.  These
conditions can differ significantly if $\lambda_2$ is close to $1$ or
$\lambda_N$ is close to $-1$.

With DGD, the communication between agents can take place in parallel
with the gradient computations, whereas this is not possible for
diffusion. However, this difference between the two algorithms is only
significant if the time spent on communication and the total time
spent on computation (counting all local gradient updates, see
Section~\ref{sec:DGD}.\ref{subsec:multilocal}) are of comparable
order.

\subsubsection*{The Special Case of a Complete Graph}

For the special case of a complete graph, $\bW=\bone\bone^\T/N$. Then
${\lambda_2=\cdots=\lambda_N=0}$, and ${(\bI-\bW)^\dagger\bW=\bzero}$.

\begin{itemize}
\item  For DGD, \eqref{eq:normxGxs} specializes to
  \begin{align}
    \norm{\hat \bx - \bx_*} & \le \eta \frac{L}{\mu}   \norm{\nafs}   .
    \label{eq:dgdfinalSc} 
  \end{align}
  
\item For diffusion, ${\Lambda = 0}$ so \eqref{eq:normxGxs} holds with
  equality. In this case the fixed point equals the optimal point:
  ${\hat\bx=\bx_*}$. This can also be seen by setting $\bZ=\bW$ in
  \eqref{eq:fixedpunif1}:
  \begin{align}
    \bx_* \eqa \bW\bx_* \eqb \bW \bx_* - \eta \bW\nabla f(\bx_*),
  \end{align}
  where in (a) we used \eqref{eq:xstarnull} and in (b) we used
  \eqref{eq:bonenafs}. 
  
\item The step size condition for DGD, \eqref{eq:eta2}, becomes
  ${\eta\le 1/(L+\mu)}$ -- to be compared with ${\eta \le 2/(L+\mu)}$
  for diffusion. (The condition ${\eta\le 1 /(L \Lambda )}$ vanishes.)
  
\end{itemize}

\begin{table*}[t!]
  \centering
  \begin{tabular}{|Sc|Sc|Sc|Sc|Sc|}
    \hline
    \textbf{Algorithm} & DGD, $T=1$ & diffusion, $T=1$ & DGD, $T>1$ & diffusion, $T>1$  \\
    \hline 
  $\norm{\bx_t-\hat\bx}$ & \eqref{eq:fixedpunif2} & \eqref{eq:fixedpunif2} & \eqref{eq:contractT1}  & \eqref{eq:contractT1} \\
  $\norm{\hat \bx-\bx_*}$ & \eqref{eq:normxGxs} and \eqref{eq:defLambda} & \eqref{eq:normxGxs} and \eqref{eq:defLambda} &  \eqref{eq:dgdfinal} & \eqref{eq:diffusionfinal} \\
 requirement on $\eta$  & \eqref{eq:eta2}  & \eqref{eq:etacondition} and  ${\eta\le 1 /(L \Lambda )}$ & \eqref{eq:eta2}  & \eqref{eq:etacondition} \\
    \hline
  \end{tabular}
  \vspace*{3mm}
  \caption{Summary of the main convergence bounds for DGD and diffusion.}
  \label{tab:summary}
  \hrule 
\end{table*}
  
\subsection{Multiple Local Gradient Updates}\label{subsec:multilocal}

The agents may perform multiple, say $T$, local gradient updates
before averaging the iterates with their neighbors.  We next extend
the analyses to this case.

Let $\bx_{t,i}$ be a vector whose $n$th component denotes the iterate
at agent $n$ at time $t$, after the $i$th local gradient update. Using
these auxiliary variables, we can describe DGD and diffusion with $T$
local gradient updates as follows:
\begin{align}
    \bx_{t,0} & = \bx_t , \label{eq:xt1Ga0}  \\
    \bx_{t,1} & = \bx_{t,0} - \eta \nabla f(\bx_{t,0}) \label{eq:xt1Ga1} , \\
    \bx_{t,2} & = \bx_{t,1} - \eta \nabla f(\bx_{t,1}) \label{eq:xt1Ga2} , \\
    \cdots  & \\
    \bx_{t,T-1} & = \bx_{t,T-2} - \eta \nabla f(\bx_{t,T-2}) \label{eq:xt1GaT1} , \\
    \bx_{t+1} & = \bW\bx_{t,T-1} - \eta \bZ \nabla f(\bx_{t,T-1}) . \label{eq:xt1GaT} 
\end{align}
Note that DGD and diffusion only differ in the last step, where the
final local gradient update is performed.

\subsubsection{Convergence to a Fixed Point $\hat\bx$}

We can write \eqref{eq:xt1Ga0}--\eqref{eq:xt1GaT} as
\begin{align}
  \bx_{t+1} = \varphi(\bx_t),  \label{eq:DGDfixm}
\end{align}
where ${ \varphi(\bx)=\phi_T(\cdots(\phi_1(\bx)))}$, and the mappings
$\{\phi_1(\cdot),\ldots,\phi_T(\cdot)\}$ are defined through
\begin{align}
  \phi_i(\bx) & = \bx - \eta \nabla f(\bx), \  i=1,\ldots,T-1  \label{eq:defphii1} \\
  \phi_T(\bx) & =  \bW\bx - \eta\bZ\nabla f(\bx). \label{eq:defphii2}
\end{align}

Under suitable step size conditions, ${\varphi(\cdot)}$ in
\eqref{eq:DGDfixm} is a contraction.  To show this, note that by the
calculation in Section~\ref{sec:GD},
$\{\phi_1(\cdot),\ldots,\phi_{T-1}(\cdot)\}$ are contractions, with
factor ${1-\eta\mu}$.  Next, for DGD, a calculation identical to
\eqref{eq:DGDfix2} shows that if \eqref{eq:eta2} holds,
$\phi_T(\cdot)$ is a contraction with factor ${1- \eta\mu}$. For
diffusion, the calculation in \eqref{eq:diffusioncontractive1} shows
that if \eqref{eq:etacondition} holds, $\phi_T(\cdot)$ is a
contraction with factor ${1- \eta\mu}$.  The C$^3$T then shows that
${\varphi(\cdot)}$ is a contraction, with factor $ (1-\eta\mu)^T$.
Hence, $\bx_t$ converges to a fixed point, $\hat\bx$, which satisfies
${\hat\bx=\varphi(\hat\bx)}$. Also,
\begin{align}
  \norm{\bx_t - \hat\bx} \le  (1-\eta\mu)^{Tt} \norm{ \bx_0 - \hat \bx}.  \label{eq:contractT1}
\end{align}

\subsubsection{Distance from $\hat\bx$ to $\bx_*$}

The bounding of ${\norm{ \hat\bx-\bx_*}}$ is complicated by the
nonlinear nature of the nested mappings in
\eqref{eq:xt1Ga0}--\eqref{eq:xt1GaT}.  In
Appendix~\ref{app:multilocal} we obtain the bounds \eqref{eq:dgdfinal}
for DGD, and \eqref{eq:diffusionfinal} for diffusion, shown on top of
the next page.
\begin{figure*}
 \begin{align}
   \mbox{DGD:} \quad   \norm{\hat \bx - \bx_*}
   & \le \eta \ppb{ \frac{T-1}{2}\frac{L}{\mu} + \pp{1+\frac{L}{\mu}}\pp{(T-1) \mnorm{(\bI-\bW)^\dagger\bW} + \frac{1}{1-\lambda_2}}} \norm{\nafs}
   + \O(\eta^2T^2)   \label{eq:dgdfinal}  \\
   \mbox{diffusion:} \quad    \norm{\hat \bx - \bx_*}
   & \le \eta \ppb{  \frac{T-1}{2}\frac{L}{\mu} + T\pp{1+\frac{L}{\mu}} \mnorm{(\bI-\bW)^\dagger\bW}} \norm{\nafs}
   + \O(\eta^2 T^2)  \label{eq:diffusionfinal}
 \end{align}
 \hrule
\end{figure*}

Note from \eqref{eq:dgdfinal}--\eqref{eq:diffusionfinal} that the
bound on ${\norm{\hat \bx - \bx_*}}$ increases linearly with
increasing $T$.  With additional local gradient updates, the iterates
at different agents diverge from each other and move towards the
respective minimizers of the local objectives $\{ f_n(\cdot) \}$. The
only remedy is to reduce $\eta$.

\subsubsection{The Special Case of a Federated Learning Setup}

In federated learning, after every iteration each agent transmits its
iterate to a central server, whereafter the server averages these
iterates and broadcasts the average back to the agents.  Typically,
each agent performs $T>1$   local gradient updates in each
iteration. This results in the algorithm
\begin{align}
  \bx_{t,0} & =  \bx_t  ,  \label{eq:diffusion1aF}  \\
  \bx_{t,i+1} & = \bx_{t,i}  - \eta\nabla f(  \bx_{t,i})  ,  \quad  i=0,\ldots,T-1,   \label{eq:diffusion1bF}\\
  \bx_{t} & = \frac{1}{N}\bone\bone^\T \cdot \bx_{t,T} .  \label{eq:diffusion1cF} 
\end{align}
Equations~\eqref{eq:diffusion1aF}--\eqref{eq:diffusion1cF} are
equivalent to \eqref{eq:xt1Ga0}--\eqref{eq:xt1GaT} with
${\bZ=\bW=\bone\bone^\T/N}$ (and using an auxiliary variable
$\bx_{t,T}$).  Thus federated learning is an instance of diffusion
on a complete-graph topology, with multiple local gradient
updates. Hence,  \eqref{eq:diffusionfinal} applies.  Since
${(\bI-\bW)^\dagger\bW=\bzero}$, \eqref{eq:diffusionfinal} reduces to
\begin{align}
  \norm{\hat \bx - \bx_*} & \le \eta { \frac{T-1}{2}\frac{L}{\mu}}
  \norm{\nafs} + \O(\eta^2T^2).  \label{eq:fedlearning}
\end{align}

\subsubsection{The Special Case of $T=1$}

In the special case of a single local gradient update ($T=1$), 
\eqref{eq:DGDfixm} reduces to \eqref{eq:DGDfix}, and
the bounds \eqref{eq:dgdfinal} and \eqref{eq:diffusionfinal} reduce to
\begin{align}
  \hspace*{-5mm}  \norm{\hat \bx - \bx_*}
  & \le \eta \pp{1+\frac{L}{\mu}}  \frac{1}{1-\lambda_2} \norm{\nafs} + \O(\eta^2)     \label{eq:dgdfinalS}
\end{align}
for DGD, and
\begin{align}
  \norm{\hat \bx - \bx_*} &  \le \eta \pp{1+\frac{L}{\mu}}  \mnorm{(\bI-\bW)^\dagger\bW} \norm{\nafs} \\
  & \qquad + \O(\eta^2)   \label{eq:diffusionfinalS}
\end{align}
for diffusion.  Equations
\eqref{eq:dgdfinalS}--\eqref{eq:diffusionfinalS} may be compared to
\eqref{eq:normxGxs}, which was obtained using a different approach.

\subsection{Summary and Comments}

Table~\ref{tab:summary} summarizes the bounds obtained so far.
 
Many alternative analysis approaches exist.  For example, as shown in
\cite{yuan2016convergence}, the DGD iteration \eqref{eq:stdDGD} is
equivalent to GD on the Lyapunov function ${ f(\bx ) + \bx^\T
  (\bI-\bW) \bx /(2\eta) }$.  However, for the diffusion iteration
\eqref{eq:diffusionatc}, no such Lyapunov equivalent exists to our
knowledge.

For DGD, comparing our bound on ${\norm{\bx_t-\bx_*}}$ with
\cite[Cor.~9]{yuan2016convergence}, one can show that if $\eta$
satisfies \eqref{eq:eta2}, our contraction speed is faster than the
linear convergence factor in \cite{yuan2016convergence}; that is, ${
  c_3\ge 1-\eta\mu }$, with $c_3$ as in \cite{yuan2016convergence}.
One can also show that for values of $\eta$ permitted by
\cite[Cor.~9]{yuan2016convergence}, the bound on the asymptotic error
therein (the second plus third term in
\cite[Cor.~9]{yuan2016convergence}) is larger than our bound on
${\norm{ \hat\bx - \bx_*}}$ in \eqref{eq:normxGxs}.

\subsection{Time-Varying Step Size}\label{sec:addext}

The analyses extend directly to the case when the step size depends on
the iteration index.  We limit the discussion here to the case of a
single local gradient update.

Let $\eta_t$ be the step size used in the $t$th iteration, and let
$\hat\bx_t$ be the fixed point associated with step size $\eta_t$;
this is the $\hat\bx$ that satisfies \eqref{eq:fixedpunif1} with
$\eta$ replaced by $\eta_t$.  The iteration \eqref{eq:unifiter} now
``chases a moving target'' (a moving fixed point).  Note that the
fixed point changes from one iteration to the next, but it is a fixed
point in the sense of the limit of $\bx_t$ if \eqref{eq:unifiter} were
iterated with the same step size.

We will need an upper bound on how much the fixed point moves between
two consecutive iterations, $\norm{ \hat\bx_{t+1} - \hat\bx_{t} }$. To
this end, note from \eqref{eq:fixedpunif1} that
\begin{align}
  \bzero & = \bZ\nabla f(\hat\bx_{t+1}) + \frac{1}{\eta_{t+1}}(\bI-\bW)\hat\bx_{t+1},    \label{eq:nablaGt11} \\
  \mbox{and} \qquad  \bzero  & =  \bZ \nabla f(\hat\bx_t) + \frac{1}{\eta_t}(\bI-\bW)\hat\bx_{t}.
  \label{eq:nablaGt1}
\end{align}
Applying the MHT to $f(\cdot)$ in \eqref{eq:nablaGt1} gives
\begin{align}
  \bzero  & = \bZ\nabla f(\hat\bx_{t+1}) + \bZ\bA_{\hat\bx_{{t+1}}\hat\bx_{t}}  (\hat\bx_{t} - \hat\bx_{{t+1}})  + \frac{1}{\eta_t}(\bI-\bW)\hat\bx_{t}   \\[-5mm]
  & \eqa \frac{-1}{\eta_{t+1}} (\bI-\bW) \hat\bx_{t+1} 
  + \bZ\bA_{\hat\bx_{{t+1}}\hat\bx_{t}}(\hat\bx_{t}- \hat\bx_{t+1}) \\ & \qquad + \frac{1}{\eta_t} (\bI-\bW)\hat\bx_{t} \\
  & = \ppb{ \bZ\bA_{\hat\bx_{{t+1}}\hat\bx_{t}} + \frac{1}{\eta_{t+1}} (\bI-\bW) } (\hat\bx_{t} - \hat\bx_{t+1})  \\
  &\qquad  +  \pp{\frac{1}{\eta_t} - \frac{1}{\eta_{t+1}}  } (\bI-\bW)\hat\bx_{t}
\end{align}
\begin{align}
  & \eqb \ppb{ \bZ\bA_{\hat\bx_{{t+1}}\hat\bx_{t}} + \frac{1}{\eta_{t+1}} (\bI-\bW) } (\hat\bx_{t}- \hat\bx_{t+1}) \\
  & \qquad -\eta_t\pp{\frac{1}{\eta_t} - \frac{1}{\eta_{t+1}}  }\bZ \nabla f(\hat\bx_t),
  \label{eq:nablaGt2}
\end{align}
where in (a) we  used \eqref{eq:nablaGt11}, and in (b) we used
again \eqref{eq:nablaGt1}.

Now apply the distance bounding result in
Appendix~\ref{subsubsec:auxbnd} with ${\bd=\hat\bx_{t} -
  \hat\bx_{t+1}}$ and
\begin{align}
  \bv = -\eta_t\pp{\frac{1}{\eta_t} - \frac{1}{\eta_{t+1}}  }\nabla f(\hat\bx_t),
\end{align}
noting that ${\Piu\nabla f(\hat\bx_t)=\bzero}$. We obtain, under
appropriate conditions on $\eta_t$:
\begin{align}
  \norm{ \hat\bx_{t+1} - \hat\bx_{t} }  
  & \le \left| \frac{1}{\eta_t} - \frac{1}{\eta_{t+1}} \right|\eta_{t+1} \eta_t \frac{L}{\mu}  \Lambda\norm{\nabla f(\hat\bx_{t})} \\
  & \lea \left| \frac{1}{\eta_t} - \frac{1}{\eta_{t+1}} \right| \eta_{t+1}\eta_t\frac{L}{\mu}   \Lambda \cdot
  2 \norm{ \nafs} \\
  & = 2{| \eta_t  -  \eta_{t+1}  |} \frac{ L}{\mu}\Lambda       \norm{ \nafs} ,  \label{eq:xgt1gt}
\end{align}
where in (a) we used that 
\begin{align}
  \norm{\nabla f(\hat\bx_{t})} &  =  \norm{ \nafs + \bA_{\bx_*\hat\bx_t}(\hat\bx_{t} - \bx_*) } \\
  & \le \norm{ \nafs} + \norm{\bA_{\bx_*\hat\bx_t}(\hat\bx_{t} - \bx_*)} \\
  & \le \norm{ \nafs} + L \norm{\hat\bx_{t} - \bx_*} \\
  & \lea  \norm{ \nafs} + \eta_t L \frac{L}{\mu} \Lambda  \norm{ \nafs }  \\
  & \le 2 \norm{ \nafs} . \label{eq:boundxgt}
\end{align}
In \eqref{eq:boundxgt}, (a) is a consequence of \eqref{eq:normxGxs}
for small enough $\eta_t$.

With a bound on the fixed point drift over time, we bound the distance
between the iterates and the fixed points as
\begin{align}
  & \hspace*{-5mm} \quad  \norm{ \bx_{t+1} - \hat\bx_{t+1} } 
  \lea   \norm{ \bx_{t+1} - \hat\bx_{t} }  + \norm{ \hat\bx_{t} - \hat\bx_{t+1}   }  \\
  & \hspace*{-5mm} \le (1-\eta_t \mu)  \norm{ \bx_t - \hat\bx_{t} }  + \norm{ \hat\bx_{t} - \hat\bx_{t+1}   }  \\
  & \hspace*{-5mm} \leb (1-\eta_t \mu)  \norm{ \bx_t - \hat\bx_{t} }  +   2| \eta_t  -  \eta_{t+1}  |  \frac{L}{\mu} \Lambda    \norm{ \nafs} ,
  \label{eq:contrvareta}
\end{align}
where (a) follows from the triangle inequality, and (b) from
\eqref{eq:xgt1gt}.  By induction over $t$, and using that ${
  \norm{\bx_0-\hat\bx_0} \le \norm{\bx_0 - \bx_*} + \norm{\hat\bx_0
    -\bx_* } }$, we obtain \eqref{eq:inductionvareta1}, on the next
page, where ${ \norm{\hat\bx_0 -\bx_* } }$ is bounded by
\eqref{eq:normxGxs}.  Finally, we bound the total error as
\begin{align}
  \norm{   \bx_{t}  - \bx_* } & \lea   \norm{   \bx_{t}  - \hat\bx_{t} }  +   \norm{ \hat\bx_{t} - \bx_* } \\
  & \leb  \norm{   \bx_{t}  - \hat\bx_{t} }  +  \eta_t \frac{L}{\mu} \Lambda \norm{\nafs},
\end{align}
where (a) is the triangle inequality, (b) follows from
\eqref{eq:normxGxs}, and ${ \norm{ \bx_{t} - \hat\bx_{t} } }$ is
upper-bounded by \eqref{eq:inductionvareta1}.
\begin{figure*}
  \begin{align}
     \norm{   \bx_{t}  - \hat\bx_{t} }   & \le \pp{ \prod_{i=0}^{t-1} (1-\eta_i \mu) } \pp{\norm{\bx_0 - \bx_*} + \norm{\hat\bx_0 -\bx_* }}
       +    2\frac{L}{\mu} \Lambda \sum_{i=0}^{t-1} | \eta_i  -  \eta_{i+1}  | \prod_{j=i+1}^{t-1} (1-\eta_j \mu)
    \cdot    \norm{ \nafs} 
    \label{eq:inductionvareta1}
  \end{align}
  \hrule
\end{figure*}

\subsubsection*{Decreasing the Step Size as $\sim 1/t$}

Based on \eqref{eq:inductionvareta1}, the sequence of step sizes, $\{
\eta_t \}$, may be designed to achieve some desired speed of
convergence. An important example is when $\eta_t$ decreases with $t$
as
\begin{align}
  \eta_t = \frac{\eta_0}{t/\tau+1},
\end{align}
for some positive constants $\eta_0$ and $\tau$ that determine the
initial step size and the decay rate, respectively.  In this case, the
total error can be made to vanish as $\O(1/t)$.

Suppose $\eta_0$ is small enough that the calculations leading to
\eqref{eq:inductionvareta1} hold.  Then for any non-negative $i$,
\begin{align}
  & \qquad  \prod_{j=i}^{t-1} (1-\eta_j\mu)  = \exp\ppb{ \sum_{j=i}^{t-1} \ln\pp{1-\frac{\eta_0\mu}{j/\tau+1}}  }  \\
  & \le  \exp\ppb{ -\eta_0\mu\tau \sum_{j=i}^{t-1} \frac{1}{j+\tau}   } \\
  & \le  \exp\ppb{ -\eta_0\mu\tau \int_{i+\tau}^{t+\tau} dz \frac{1}{z}  } \\
  & =  \exp\ppb{ -\eta_0\mu\tau \ln\pp{\frac{t+\tau}{i+\tau}}  } 
  = \pp{\frac{i+\tau}{t+\tau}}^{\eta_0\mu\tau}. \label{eq:prodetajmu} 
\end{align}
The first term of \eqref{eq:inductionvareta1} can be upper-bounded by
using \eqref{eq:prodetajmu} with $i=0$:
\begin{align}
  \prod_{i=0}^{t-1} (1-\eta_i\mu) &  \le \pp{\frac{\tau}{t+\tau}}^{\eta_0\mu\tau}.  \label{eq:firsttermbnd}
\end{align}
The second term of \eqref{eq:inductionvareta1} can be upper-bounded, for ${\tau \neq 1/(\eta_0\mu)}$, as
follows: 
\begin{align}
  & \qquad  \sum_{i=0}^{t-1} (\eta_i - \eta_{i+1}) \prod_{j=i+1}^{t-1} (1-\eta_j\mu) \\
  & \lea \eta_0\tau \sum_{i=0}^{t-1} \frac{1}{(i+\tau)(i+\tau+1)} \pp{\frac{i+1+\tau}{t+\tau}}^{\eta_0\mu\tau} \\
  & \le \frac{ \eta_0\tau(1+1/\tau)}{(t+\tau)^{\eta_0\mu\tau} }  \sum_{i=0}^{t-1} (i+1+\tau)^{\eta_0\mu\tau-2} \\
  & = \frac{ \eta_0(\tau+1)}{(t+\tau)^{\eta_0\mu\tau} }  \sum_{i=1}^{t} (i+\tau)^{\eta_0\mu\tau-2} \\
  & \leb \frac{ \eta_0(\tau+1)}{(t+\tau)^{\eta_0\mu\tau} }  \int_{0}^{t+1} dz\ (z+\tau)^{\eta_0\mu\tau-2} \\
  & = \frac{ \eta_0(\tau+1)}{(t+\tau)^{\eta_0\mu\tau} }  \frac{ (t+\tau+1)^{\eta_0\mu\tau-1} - \tau^{\eta_0\mu\tau-1}}{\eta_0\mu\tau-1}, \label{eq:sectermbnd}
\end{align}
where in (a) we used \eqref{eq:prodetajmu}. When upper-bounding the
sum by an integral in (b), note that depending on $\eta_0\mu\tau$, the
terms in the sum may either be increasing or decreasing with $i$.
Equations \eqref{eq:firsttermbnd} and \eqref{eq:sectermbnd} together
show that if ${\tau > 1/(\eta_0\mu)}$, \eqref{eq:inductionvareta1}
approaches zero as $\O(1/t)$: the first term is
$\O(1/t^{\eta_0\mu\tau})$ and the second term is $\O(1/t)$.  If ${\tau
  < 1/(\eta_0\mu)}$, \eqref{eq:inductionvareta1} approaches zero as
$\O(1/t^{\eta_0\mu\tau})$; in this case, both terms are of that order.
At the ``critical point'' when ${\tau = 1/(\eta_0\mu)}$, the last step
in \eqref{eq:sectermbnd} does not hold but by evaluating the integral
one can show that \eqref{eq:inductionvareta1} decays as
$\O(\log(t)/t)$.

A similar approach to the analysis of DGD with time-varying step sizes
was used in \cite{Michelusi-NCOTADGD}.

\section{Noisy DGD and Diffusion}\label{sec:noisygrad}

The DGD and diffusion iterations in \eqref{eq:unifiter}, and their
counterparts with multiple local gradient updates
[\eqref{eq:xt1Ga0}--\eqref{eq:xt1GaT}] may be noisy.  There are
several possible origins of the noise: noise in the gradients due to
random subsampling of data (stochastic DGD and diffusion);
communication noise because of quantization or transmission errors
when gradients are communicated between agents; or as an instance of
the latter, randomness in the topology owing to, for example,
communication link failures. Because of the noise, the iterates
$\{\bx_t\}$ become random variables.  Next we analyze noisy DGD and
diffusion in the MHT+contraction framework.

\subsection{Data Subsampling: Stochastic DGD and Diffusion}\label{subsec:sdgd}

The effect of random data subsampling can be modeled by adding noise
onto the gradients. We consider first the case of a single local
gradient update.  Equation~\eqref{eq:unifiter} becomes
\begin{align}
  \bx_{t+1} & = \bW\bx_{t} - \eta \bZ(\nabla f(\bx_t) + \beps_t),  \label{eq:samplingnoise}
\end{align}
where $\beps_{t}$ is the gradient noise at iteration $t$. Note that
this noise is scaled by the step size, $\eta$.
 
The noise $\beps_t$ is statistically dependent on $\bx_t$, and
potentially on other past iterates.  However, if in every iteration
the data are subsampled uniformly at random, $\beps_{t}$ will have
zero mean conditionally on $\bx_t$,
\begin{align}
  \E{\beps_{t}  | \bx_t   } & =\bzero  .  \label{eq:noiseass1} 
\end{align}
We will assume that  the conditional noise variance is at most affine in
the norm-square of the iterates, so that
\begin{align}
\hspace*{-10mm}   \E{ \snorm{\eta\bZ\beps_{t}} | \bx_t } \le    \E{ \snorm{\eta\beps_{t}} | \bx_t } \le \eta^2 \pp{\sigma +  \omega \norm{\bx_t} }^2 , \label{eq:conditionalvar}
\end{align}
for some non-negative constants $\sigma$ and $\omega$.
Equation~\eqref{eq:conditionalvar} is a standard model for sampling
noise in stochastic gradient algorithms; for example, it is implied by
\cite[Eq.~(48)]{chen2012diffusion} (see the remark at the end of
Section~\ref{sec:tools}.\ref{sec:noisycontr}).\footnote{\setstretch{0.7}As
pointed out in \cite{chen2012diffusion}, this condition is similar to
\cite[Eq.~(4.3)]{bertsekas2000gradient}. While
\eqref{eq:conditionalvar} and \cite[Eq.~(48)]{chen2012diffusion}
express the upper bound on the conditional mean-square noise in terms
of $\norm{\bx_t}$, \cite[Eq.~(4.3)]{bertsekas2000gradient} instead
expresses this upper bound in terms of $\norm{\nabla f(\bx_t)}$; note
that $\norm{\bx_t}$ and $\norm{\nabla f(\bx_t)}$ can be related to one
another using the MHT.}

Suppose $\eta$ satisfies the condition (from the NC$^2$T)
\begin{align}
  (1-\eta\mu)^2 + \eta^2\omega^2 < 1 . \label{eq:etacondnoise}
\end{align}
Applying the NC$^2$T to the contractions defined in
Section~\ref{sec:DGD}.\ref{sec:convfix} with ${c=1-\eta\mu}$, and
$\sigma$ and $\omega$ scaled by $\eta$, gives
\begin{align}
  &   \sqrt{ \E{ \snorm{\bx_t - \hat\bx}} }  
  \\  & \hspace*{-10mm} \le \hat d + \sqrt{(1-\eta\mu)^2 + \eta^2\omega^2}^{\,t} \pp{  {  \norm{\bx_0 - \hat\bx}} - \hat d}^+,
  \label{eq:NC3Tappl1}
\end{align}
where
\begin{align}
  \hat d & =  \frac{\eta\omega\norm{\hat\bx}  + \eta\sigma}{\sqrt{1-(1-\eta\mu)^2}-\eta\omega} 
  \lea \eta \frac{\omega \norm{\hat\bx - \bx_*}+ \omega\norm{\bx_*}  + \sigma}{\sqrt{2 \eta\mu - \eta^2\mu^2} -\eta\omega }  ;
  \label{eq:dhatnoisygrad}
\end{align}
in (a) we used the triangle inequality.
For small enough $\eta$, 
\begin{align}
  \frac{\eta}{ \sqrt{2 \eta\mu - \eta^2\mu^2} -\eta\omega  } \le \frac{\sqrt{\eta}}{\sqrt{\mu}},  \label{eq:smalleneta}
\end{align}
which gives, from \eqref{eq:dhatnoisygrad}, the following simpler expression:
\begin{align}
  \hat d  & \le \sqrt{\eta} \frac{\omega \norm{\hat\bx - \bx_*}+ \omega\norm{\bx_*}  + \sigma}{\sqrt{  \mu} } .  \label{eq:dhatsamplnoise1}
\end{align}

To bound the total error, we use \eqref{finalbound}:
\begin{align}
  &  \sqrt{\E{\snorm{\bx_{t} - \bx_*}}   } 
  \le \hat d   +   \norm{ \hat\bx - \bx_*   } + \sqrt{(1-\eta\mu)^2 + \eta^2\omega^2}^{\,t} \\[-5mm]  
  & \qquad \cdot \pp{   { \norm{\bx_0 - \bx_*} } + \norm{\bxhs} - \hat d}^+.
  \label{eq:noisygradfinalerror1}
\end{align}
The distance $\norm{ \hat\bx - \bx_* }$ that enters into
\eqref{eq:noisygradfinalerror1}, both directly and via $\hat d$, is
obtained from \eqref{eq:normxGxs}.  Consider the three terms in
\eqref{eq:noisygradfinalerror1}.  If $\eta$ satisfies
\eqref{eq:smalleneta} then the contraction factor satisfies
\begin{align}
 \sqrt{(1-\eta\mu)^2 + \eta^2\omega^2}  \leq \sqrt{1-\eta\mu} \leq e^{-\eta\mu/2} , \label{eq:etacondnoise2}
\end{align}
and \eqref{eq:etacondnoise} in particular.  For given $\epsilon>0$,
there exists $\eta$ such that the first two terms in
\eqref{eq:noisygradfinalerror1} sum up to less than $\epsilon/2$. For
that $\eta$, there is a value of $t$ such that the third term of
\eqref{eq:noisygradfinalerror1} is less than $\epsilon/2$.  In
particular, $\eta$ should be of order ${\sim {\epsilon}^2 }$ to make
the first two terms small, and $t$ should be of order ${\sim
  \log(1/\epsilon)/\epsilon^2 }$ to make the third term small.

\subsubsection{The Impact of the Noise}

In \eqref{eq:noisygradfinalerror1}, ${\norm{\bxhs}=\O(\eta)}$, whereas
${\hat d=\O(\sqrt{\eta})}$.  Hence, for small $\eta$,
${\sqrt{\E{\snorm{ \bx_t - \bx_* }}}}$ is dominated by the effects of
the gradient noisiness.

The noise has two effects. First, if $\omega>0$, it slows down the
convergence to the fixed point, $\hat\bx$, by increasing the
contraction factor from ${1-\eta\mu}$ to ${\sqrt{(1-\eta\mu)^2 +
    \eta^2\omega^2}}$, requiring a correspondingly smaller $\eta$.
Second, the noise increases $\hat d$, by an amount dependent on both
$\omega$ and $\sigma$.  Note, however, that the fixed point itself is
unaffected by the presence of noise.

\subsubsection{Multiple Local Gradient Updates}

To analyze convergence with multiple local gradient updates, we apply
the NC$^3$T to the contractions
$\{\phi(\cdot)\}$ in \eqref{eq:defphii1}--\eqref{eq:defphii2}.  We
obtain a bound with similar appearance as that in
\eqref{eq:noisygradfinalerror1}:
\begin{align}
  &    \sqrt{\E{\snorm{\bx_{t} - \bx_*}}   } 
  \le \hat d   +   \norm{ \hat\bx - \bx_*   } + \sqrt{(1-\eta\mu)^2 + \eta^2\omega^2}^{\,tT} \\[-5mm]  
  & \qquad \cdot \pp{ {   \norm{\bx_0 - \bx_*}} + \norm{\bxhs} - \hat d}^+ .
  \label{eq:noisygradfinalerror1M}
\end{align}
For small enough $\eta T$,
\begin{align} 
  \hat d & \le \frac{\eta\omega M + \eta\sigma}{     \sqrt{1-(1-\eta\mu)^2}-\eta\omega } 
  \lea \sqrt{\eta} \frac{\omega \norm{\bx_*} + \sigma}{\sqrt{\mu}}  ,
  \label{eq:dhatnoisygradM}
\end{align}
where $M$ quantifies how much the fixed point
shifts because of the multiple local gradient updates; in (a), we used
the bound on $M$ from \eqref{eq:unibndxihat} in
Appendix~\ref{app:multilocal}:
\begin{align}
  M = \max\{ \norm{\tilde\bx_1},\ldots,\norm{\tilde\bx_T} \} \le  \norm{\bx_*}+\O(\eta T),
\end{align}
where $\{\tilde\bx_i\}$ are the auxiliary iterates defined in that
appendix.  Finally, in \eqref{eq:noisygradfinalerror1M},
$\norm{\hat\bx - \bx_*}$ is bounded by \eqref{eq:dgdfinal} for DGD and
\eqref{eq:diffusionfinal} for diffusion.

\subsection{Noisy Inter-Agent Communication}\label{subsec:commnoise}

The communication between agents may be subject to errors caused by
noise, interference, or signal quantization (e.g., sparsification).
Such errors can be modeled by including a noise term $\beps_{t}$ that
enters additively in \eqref{eq:unifiter}:
\begin{align}
  \bx_{t+1} & =  \bW\bx_{t} - \eta \bZ \nabla f(\bx_t) + \beps_t. \label{eq:xtcommnoise}
\end{align}
The analysis follows that for noisy gradients
(Section~\ref{sec:noisygrad}.\ref{subsec:sdgd}), but with an important
difference: in the noisy-gradients case, $\beps_{t}$ is scaled by
$\eta$; here, it is not, since the noise affects the iterates
communicated among the agents rather than the gradient updates.

We will assume that $\{\beps_t\}$ satisfy
\eqref{eq:noiseass1} and that
\begin{align}
  \E{   \snorm{\beps_{t}} | \bx_t }   &    \le  (\sigma + \omega \norm{\bx_t})^2, \label{eq:noisylinks2}
\end{align}
for some non-negative constants $\sigma$ and $\omega$.  Equations
\eqref{eq:noiseass1} and \eqref{eq:noisylinks2} hold for example for
noise and interference in analog transmission, and for quantization
noise, if appropriate dithering is applied.

\subsubsection{Scaling the Consensus Weights}\label{subsec:tuningspeed}

There may not exist an $\eta$ such that
${(1-\eta\mu)^2+\omega^2<1}$. Hence, the NC$^2$T is not directly
applicable; in fact, the iteration \eqref{eq:xtcommnoise} may diverge.
To guarantee convergence, we scale $\bW$ using a \emph{consensus step
size}, $\gamma$, ${0<\gamma\le 1}$.  With this modification, the
noise-free iteration \eqref{eq:unifiter}, becomes
\begin{align}
  \bx_{t+1} & = (1-\gamma) \bx_t + \gamma  \bW\bx_{t}  - \eta \bZ \nabla f(\bx_t)    \\
  & = \bW' \bx_t - \eta \bZ \nabla f(\bx_t), \label{eq:xt1GaS}
\end{align}
where 
\begin{align}
\bW'= (1-\gamma) \bI + \gamma \bW.  \label{eq:Wlift}
\end{align}
(Note that $\bW'^\T=\bW'$.)  The effect of the consensus step size
scaling is to average the iterates over time, and the role of
$1-\gamma$ is similar to the forgetting factor in an autoregressive
lowpass filter.  Compared to the nominal $\bW$, the effective weight
matrix $\bW'$ in \eqref{eq:Wlift} is pushed towards $\bI$: the
eigenvalues of $\bW'$ are ${ \{ 1 - \gamma +\gamma\lambda_n \} }$ and
they satisfy \eqref{eq:Weigs} if the eigenvalues of $\bW$,
$\{\lambda_n\}$, do.  For DGD, the requirement on the step size,
(\ref{eq:eta2}), becomes
\begin{align}
  \eta \le \frac{2-\gamma + \gamma\lambda_N}{L+\mu},
\end{align}
while the requirement for diffusion is unchanged.

With the weight scaling, the noisy iteration \eqref{eq:xtcommnoise}
becomes
\begin{align}
  \bx_{t+1} & = \bW' \bx_t - \eta\bZ\nabla f(\bx_t) + \gamma \beps_t. \label{eq:xtcommnoise2}
\end{align}
 In particular,
$\gamma$ scales the communication noise in \eqref{eq:xtcommnoise2}: by
slowing down the consensus process, the noise is averaged over time,
effectively scaling its variance by $\gamma^2$.    

\subsubsection{Bounding the Error}

Application of the NC$^2$T with $\sigma$ and $\omega$ scaled by
$\gamma$, assuming that $\gamma$ is small enough that ${(1-\eta\mu)^2
  + \gamma^2\omega^2<1}$, gives
\begin{align}
  &   \sqrt{\E{\snorm{\bx_{t} - \bx_*}}   } 
  \lea \hat d + \norm{ \hat\bx - \bx_*   } + \sqrt{(1-\eta\mu)^2 + \gamma^2\omega^2}^{\,t} \\[-5mm]  
  & \qquad \cdot \pp{  {  \norm{\bx_0 - \bx_*}} + \norm{\bxhs} - \hat d}^+  ,   \label{eq:noisygradfinalerror2} 
\end{align}
where
\begin{align}
  \hat d & = \frac{\gamma\omega\norm{\hat\bx} + \gamma\sigma}{\sqrt{1-(1-\eta\mu)^2}-\gamma\omega}   \\
  & \le \gamma\frac{\omega\norm{\bxhs}+\omega\norm{\bx_*} +  \sigma}{\sqrt{2\eta\mu - \eta^2\mu^2}-\gamma\omega}   \\
  & \lea \frac{\gamma}{\sqrt{\eta}}\frac{\omega\norm{\bxhs}+\omega\norm{\bx_*} +  \sigma}{\sqrt{ \mu}}  ,   \label{eq:dhatcommnoise}
\end{align}
where (a) holds for small enough $\eta$ and small enough
$\gamma/\sqrt{\eta}$.  The distance $\norm{ \hat\bx - \bx_* }$ that
enters in \eqref{eq:noisygradfinalerror2} (directly and via $\hat d$)
is bounded by \eqref{eq:normxGxs}, with $\bW$ replaced by $\bW'$.  In
particular, $\lambda_2$ is replaced by the second-largest eigenvalue
of $\bW'$, ${1-\gamma+\gamma\lambda_2}$; hence, ${1-\lambda_2}$ is
replaced by ${\gamma(1-\lambda_2)}$ in \eqref{eq:dgddiffcomp}.  In
view of \eqref{eq:normxGxs} and \eqref{eq:dgddiffcomp}, this means
that ${\norm{ \hat\bx - \bx_* } = \O(\eta/\gamma)}$.  Hence $\eta$ and
$\gamma$ must be jointly selected to make $ \sqrt{\E{\snorm{\bx_{t} -
      \bx_*}} } $ small.  By taking $\gamma=\eta^{3/4}$, the first and
the second terms in \eqref{eq:noisygradfinalerror2} are of the same
order, $\sim \eta^{1/4}$.  With this choice of $\gamma$, the
contraction factor satisfies
\begin{align}
  \sqrt{(1-\eta\mu)^2 + \eta^{3/2}\omega^2}\leq \sqrt{1-\eta\mu} \leq e^{-\eta\mu/2},
\end{align}
for small enough $\eta$.  For the so-selected, small enough, $\eta$
and $\gamma$, the third term of \eqref{eq:noisygradfinalerror2}  
goes to $0$ as $t\to\infty$.

A similar tuning of the learning ($\eta$) and consensus ($\gamma$)
step sizes was suggested in \cite{8786146} to mitigate quantization
noise and in \cite{Michelusi-NCOTADGD} to mitigate wireless
propagation effects.

\subsection{Random Topology}\label{sec:linkfail}

In some applications, the topology is random.  This could happen if
communication links between agents are activated (scheduled), or fail, 
at random. The  iteration \eqref{eq:unifiter} then becomes
\begin{align}
  \bx_{t+1} & =  \bQ_t \bx_{t}  - \eta \bZ_t \nabla f(\bx_t) ,
  \label{eq:linkfailiterN}
\end{align}
where $\bQ_t$ is a (random) weight matrix that defines how, in
iteration $t$, the agents aggregate the irregularly received updates
from their neighbors, and $\bZ_t$ is defined from $\bQ_t$ similar to
\eqref{eq:defZ}. 
To guarantee convergence, the consensus speed must be
tuned, resulting in the modified iteration,
\begin{align}
  \bx_{t+1} & = (1-\gamma) \bx_t   + \gamma \bQ_t \bx_{t}  - \eta  \bZ_t \nabla f(\bx_t) \\
  & = (1-\gamma) \bx_t   + \gamma \E{\bQ_t} \bx_{t}  - \eta \E{\bZ_t} \nabla f(\bx_t) + \gamma \beps_t,
  \label{eq:linkfailitera}
\end{align}
where $\beps_t$ is effective noise:
\begin{align}
  \beps_t & =   (\bQ_t-\E{\bQ_t}) \bx_t - \frac{\eta}{\gamma} (\bZ_t-\E{\bZ_t})\nabla f(\bx_t).  \label{eq:defepstlinkfail}
\end{align}
If $\bQ_t$ is statistically independent of $\{\bx_1,\ldots,\bx_t\}$
for all $t$ then by inspection of \eqref{eq:defepstlinkfail}, the
condition \eqref{eq:noiseass1} is satisfied.  If $\{\bQ_t\}$ are
identically distributed so that we can omit the subscript $t$, and
have bounded second-order moments, then \eqref{eq:noisylinks2} is
satisfied for some constant $\omega$.

Equation \eqref{eq:linkfailitera} has the exact same form as
\eqref{eq:xtcommnoise2}, with $\bW$ replaced by $\E{\bQ}$, and
consequently $\bW'$ in \eqref{eq:Wlift} replaced by ${(1-\gamma)\bI +
  \gamma\E{\bQ}}$.  Hence, the analysis in
Section~\ref{sec:noisygrad}.\ref{subsec:commnoise} applies, provided
that $\E{\bQ}$ satisfies the requirements made on $\bW$ in
Section~\ref{sec:DGD}.\ref{subsec:topology}.  Note that we can have
${\bQ^\T\neq\bQ}$ in general {(hence, the instantaneous
  topology could be asymmetric and directed)}, but
${\E{\bQ^\T}=\E{\bQ}}$ is required.

In case $\{ \bQ_t \}$ are non-i.i.d., $\beps_t$ and $\beps_{t'}$ will
become statistically dependent for $t\neq t'$, which causes
\eqref{eq:noiseass1} to be violated. An analysis of this case would
require modeling of the joint statistics of $\{ \bx_t \}$ over
multiple iterations.

\subsection{Summary and Comments}

Table~\ref{tab:summaryN} summarizes the bounds obtained for the noisy case.

\begin{table}[t!]
  \centering
  \begin{tabular}{|Sc|Sc|Sc|Sc|Sc|}
    \hline
    \textbf{Algorithm} &  $T=1$ &   $T>1$ \\
    \hline 
    $\sqrt{ \E{ \snorm{\bx_t - \bx_*}}  }$ & \eqref{eq:noisygradfinalerror1}    &  \eqref{eq:noisygradfinalerror1M} \\
    w.\ sampl.\ noise [Eq.~\eqref{eq:samplingnoise}]   & [$\hat d$ in \eqref{eq:dhatsamplnoise1}]  & [$\hat d$ in \eqref{eq:dhatnoisygradM}]  \\
    \hline 
    $\sqrt{ \E{ \snorm{\bx_t - \bx_*}}  }$ & \eqref{eq:noisygradfinalerror2}    &     \\
    w.\ comm.\ noise [Eq.~\eqref{eq:xtcommnoise2}]    &  [$\hat d$ in \eqref{eq:dhatcommnoise}]  &  \\
    \hline
    $\norm{\hat \bx - \bx_*}$  for DGD   & \eqref{eq:normxGxs} and \eqref{eq:defLambda}     &  \eqref{eq:dgdfinal} \\
    $\norm{\hat \bx - \bx_*}$  for diffusion   & \eqref{eq:normxGxs} and \eqref{eq:defLambda}     &  \eqref{eq:diffusionfinal} \\
    \hline
  \end{tabular}
  \vspace*{3mm}
  \caption{Summary of the main convergence bounds for the noisy case.}
  \label{tab:summaryN}
\end{table}

Note how our MHT+contraction framework allows the conditional noise
variance to scale with the squared magnitude of the iterates [see
  \eqref{eq:conditionalvar} and \eqref{eq:noisylinks2}].  In contrast,
most literature (for example,
\cite{xing2021federated,9515709,wei2022federated,9562522,pmlr-v130-ruan21a,stich2019,8870236}),
assumes that the gradient noise variance is \emph{uniformly bounded},
which corresponds to having $\omega=0$ in \eqref{eq:conditionalvar}
and \eqref{eq:noisylinks2}. That assumption is violated in many
problems of practical interest: for instance, as shown in
Appendix~\ref{sec:unboundednoise}, it cannot hold for linear
regression.

Examples of analyses in the literature that allow noise with the
property \eqref{eq:noisylinks2}, with $\omega>0$, are
\cite{9782148,8786146}.  Theorem~2 in \cite{8786146} also contains an
error bound for noisy DGD with ${\omega>0}$ (but ${\sigma=0}$).  This
theorem only gives a scaling behavior; in contrast, our bounds are
explicit.

\section{Application Example: DGD with Link Failures}\label{sec:linkfailapp}

As an application example, we consider standard DGD (${\bZ=\bI}$) with
inter-agent communication link failures.  (Extensions to diffusion are
possible, but require additional notation.)  The literature on
federated and decentralized learning with unreliable communications is
quite rich; see Section~\ref{sec:intro}.\ref{subsec:lit}.  Here we
illustrate how the MHT+contraction framework, and the results in
Sections~\ref{sec:noisygrad}.\ref{subsec:commnoise}--\ref{sec:linkfail},
can be directly used to analyze such setups.

Let $\bW$ be a matrix of nominal weights associated with some given
base topology, and denote by $p_{nn'}$ the probability that the
transmission from agent $n'$ to agent $n$ succeeds.  We assume that
the success probabilities are positive and symmetric:
$p_{nn'}=p_{n'n}>0$ for all $n$ and $n'$, and that all link failures
are mutually independent.  Given $\bW$, let $\bS$ be a random matrix
whose entries are independent and
\begin{align}
  S_{nn'} = \begin{cases} W_{nn'}, & \mbox{with probability $p_{nn'}$} \\
    0, & \mbox{with probability $1-p_{nn'}$} \end{cases}.
\end{align}
Replacing $\bW$ in \eqref{eq:stdDGD} with $\bS$ does not guarantee
convergence, but the use of the modified iteration
\eqref{eq:linkfailitera} does, if $\bQ$ is appropriately constructed
based on $\bS$.  Note that $\bS$ is beyond control (it is determined
by $\bW$ and $\{p_{nn'}\}$), but $\bQ$ can be \emph{constructed} based
on $\bS$, consistent with the topology defined by $\bW$ -- that is,
allowing $Q_{nn'}$ to be nonzero only if $W_{nn'}$ nonzero (for
${n\neq n'}$).  Note also that while ${\bW^\T=\bW}$, we have
${\bS^\T\neq\bS}$ in general.

\subsection{Known Link Success Probabilities $\{ p_{nn'} \}$}\label{sec:knownprobs}

If  $\{ p_{nn'} \}$ are \emph{known}, we construct
$\bQ$ as follows:
\begin{align}
  \hspace*{-7mm}  Q_{nn'} =
  \begin{cases}
    1 + S_{nn} - \sum_{n''=1}^N p_{nn''}W_{nn''}  , & n = n' \\
    S_{nn'} , & n \neq n',
  \end{cases} \label{eq:Zconstructknownp}
\end{align}  
Clearly,
\begin{align}
  \hspace*{-8mm}  \E{Q_{nn'}} =
  \begin{cases}
    1  - \sum_{n''=1, n''\neq n}^N p_{nn''}W_{nn''}  , & n = n' \\
    p_{nn'}W_{nn'} , & n \neq n'.
  \end{cases} \label{eq:EZconstructknownp}
\end{align}  
If the nominal weight matrix, $\bW$, satisfies the conditions in
Section~\ref{sec:DGD}.\ref{subsec:topology} then $\E{\bQ^\T}=\E{\bQ}$,
$\E{\bQ}\bone=\bone$, and the graph induced by $\E{\bQ}$ is
connected.

For the analysis in
Sections~\ref{sec:noisygrad}.\ref{subsec:commnoise}--\ref{sec:linkfail}
to apply, we additionally require that the eigenvalues of $\E{\bQ}$
satisfy \eqref{eq:Weigs}.  By the arguments in
Section~\ref{sec:discuss}.\ref{sec:suffcond}, a set of sufficient
conditions for this is that $\E{\bQ}$ is non-negative and not
bipartite.  These conditions in turn are satisfied if $\bW$ is
non-negative and not bipartite. (Note that, in contrast,
non-negativity of $\bW$ is not required by the main analysis in
Sections~\ref{sec:DGD} and \ref{sec:noisygrad}.)

We have that ${\bq_n-\E{\bq_n} =\bs_n-\E{\bs_n}}$, where $\bs_n^\T$ is
the $n$th row of $\bS$ and $\bq_n^\T$ is the $n$th row of $\bQ$.
Also,
\begin{align}
  \var{S_{nn'}} = p_{nn'} (1-p_{nn'}) W^2_{nn'}.  
\end{align}
The effective noise variance, computed via \eqref{eq:defepstlinkfail}
with $\bZ_t=\bI$ and $\bQ$ given by \eqref{eq:Zconstructknownp}, can
be upper-bounded as
\begin{align}
  & \qquad  \E{\snorm{\beps_t} | \bx_t } 
  =    \Ebig{  \sum_{n=1}^N \ppb {(\bq_n - \E{\bq_n})^\T \bx_t }^2 \Big| \bx_t} \\
  & \le
  \sum_{n=1}^N \E{\snorm{\bq_n  - \E{\bq_n} }} \cdot \snorm{\bx_t} \\
  & = \sum_{n=1}^N \E{\snorm{\bs_n  - \E{\bs_n} }} \cdot \snorm{\bx_t} \\
  & =  { \sum_{n=1}^N\sum_{n'=1}^N p_{nn'}(1-p_{nn'}) W_{nn'}^2 }  \cdot  \snorm{\bx_t} \\
  & \lea \frac{1}{4} \tr{ \bW^\T \bW}  \cdot  \snorm{\bx_t} \\
  & = \frac{1}{4} \sum_{n=1}^N  \lambda_n^2    \snorm{\bx_t}  
\,  \leb \, \frac{N}{4}    \snorm{\bx_t},
  \label{eq:epslinkfail}
\end{align}
where in (a) we used that $x(1-x)\le 1/4$ for all $x$, and in (b) we
used that ${\mnorm{\bW}\le 1}$. 

\subsection{Unknown Link Success Probabilities $\{ p_{nn'} \}$}

If $\{ p_{nn'} \}$ are \emph{unknown}, $\bQ$ in
\eqref{eq:Zconstructknownp} is not computable. Instead, we may
construct $\bQ$ as follows:
\begin{align}
  Q_{nn'} =
  \begin{cases} 1 + S_{nn} - \sum_{n''=1}^N S_{nn''}  , & n = n' \\
    S_{nn'} , & n \neq n'.
  \end{cases} \label{eq:Zconstructunknownp}
\end{align}  
Then \eqref{eq:EZconstructknownp} holds, and
\begin{align}
  \bq_n  - \E{\bq_n} & =    (\bI - \be_n\bone^\T)   (\bs_n  - \E{\bs_n}) .
\end{align}
This gives the noise variance,  
\begin{align}
  & \quad \E{\snorm{\beps_t} | \bx_t } 
     \le \sum_{n=1}^N \E{\snorm{\bq_n  - \E{\bq_n} }} \cdot \snorm{\bx_t} \\
  & = \sum_{n=1}^N \tr{  (\bI - \be_n\bone^\T)  \cov{\bs_n}  (\bI - \bone\be_n^\T) } \cdot \snorm{\bx_t} \\
  & \eqa 2 \sum_{n=1}^N  \pp{ \tr{\cov{\bs_n} }   - \var{S_{nn}}   }   \snorm{\bx_t}  \\  
  & \le   \frac{1}{2} \sum_{n=1}^N \sum_{n'=1}^N W_{nn'}^2   \snorm{\bx_t} 
   \le \frac{N}{2}    \snorm{\bx_t},
  \label{eq:epslinkfail2}
\end{align}
where in (a) we used that the components of $\bs_n$ are independent;
hence $\cov{\bs_n}$ is diagonal.  Note how the noise variance doubles
compared to the known-probabilities case.

\subsection{Effects of  $\{p_{nn'}\}$ on the Convergence Properties}

The link success probabilities $\{p_{nn'}\}$ affect the distance in
\eqref{eq:normxGxs}, which is determined by the second-largest
eigenvalue of $\E{\bQ}$ -- along with, among others, the consensus
step size $\gamma$.  If $\bW$ is non-negative (see the remark at the
end of Section~\ref{sec:linkfailapp}.\ref{sec:knownprobs}), a decrease
in any of the link success probabilities pushes this eigenvalue
towards $1$.  To see this, suppose we decrease $p_{ij}=p_{ji}$ by some
positive $\delta$, for some $(i,j)$, $i\neq j$. Let
$\overline{\E{\bQ}}$ be $\E{\bQ}$ after this decrease. This changes
$\E{Q_{ij}}$ and $\E{Q_{ji}}$ by $\delta W_{ij}$, and $\E{Q_{ii}}$ and
$\E{Q_{jj}}$ by ${-\delta W_{ij}}$. Hence,
\begin{align}
    \overline{\E{\bQ}} - \E{\bQ} 
  & =  \delta W_{ij} (\be_i\be_i^\T + \be_j\be_j^\T - \be_i\be_j^\T - \be_j\be_i^\T) \\
  & = \delta W_{ij} (\be_i-\be_j)(\be_i-\be_j)^\T \succcurlyeq \bzero,
\end{align}
so the eigenvalues of $\E{\bQ}$ increase (or stay unchanged).  This
effectively slows down the consensus speed.

The probabilities $\{p_{nn'}\}$ also affect the noise variance
${\E{\snorm{\beps_t} | \bx_t }}$. While the final upper bounds on this
variance in \eqref{eq:epslinkfail} and \eqref{eq:epslinkfail2} are not
tight enough to show this dependence, inspection of the intermediate
steps in \eqref{eq:epslinkfail} and \eqref{eq:epslinkfail2} reveals
that the variance increases if some $p_{ij}$ is decreased, as long as
$1/2 \le p_{ij} \le 1$.  

\section{Additional Discussion, Insights, and Extensions}\label{sec:discuss}

\subsection{Vector-Valued Parameters}
 
The MHT+contraction analysis approach extends directly to the case of
vector-valued parameters.  Nothing structurally changes in the problem
and it is straightforward (but notationally cumbersome) to describe
the DGD and diffusion iterations for the vector-valued case using a
single equation.  Let $d$ be the dimension of the parameter vector,
let $\bx_{t,n}$ be the $d$-dimensional iterate held by agent $n$ at
iteration $t$, and define the $N\times d$-matrix ${\bX_t=[\bx_{t,1}
    \ \cdots \ \bx_{t,N}]^\T}$. The local objectives become mappings
${f_n:\realR^d\to \realR}$, the global objective becomes a mapping
${F:\realR^d\to \realR}$, and the component-wise objective becomes a
mapping ${f: \realR^{N\times d} \to \realR }$.  The counterpart of
\eqref{eq:unifiter} becomes
\begin{align}
  \bX_{t+1} &  = \bW\bX_t -\eta  \bZ
  \begin{bmatrix} 
    \ppb{\nabla f_1(\bx_{t,1})}^\T \\
    \vdots \\
    \ppb{\nabla f_N(\bx_{t,N})}^\T 
  \end{bmatrix}.
  \label{eq:vectoreq1}
\end{align}
(Similar equations can be written up for the case with multiple local
iterations.)  Note how $\bW$ acts on each column of $\bX_t$ (data
dimension) independently.  To write \eqref{eq:vectoreq1} as a vector
iteration, formally similar to \eqref{eq:unifiter}, one may vectorize
$\bX_{t}$.  For example, the first term of \eqref{eq:vectoreq1} can be
written as
\begin{align}
  \vec{\bW\bX_t} =   \pp{ \bI\otimes \bW } \vec{\bX_t}.
\end{align}
Note that $ \bI\otimes \bW$ has the same eigenvalue spectrum as $\bW$,
but with every eigenvalue repeated $d$ times.

\subsection{Data Heterogeneity Metrics}\label{sec:hetero}

In all bounds on ${\norm{\hat\bx-\bx_*}}$, $\norm{\nafs}$ quantifies
the heterogeneity among the objectives (data) held by the different
agents.  There are, at least, two alternative metrics that can be used
to quantify this heterogeneity: the distance ${\norm{\bx_*-\bxl}}$,
and the objective function gap ${f(\bx_*)-f(\bxl)}$, where
$\bxl=[x_1^*,\ldots,x_N^*]^\T$ and $x_n^*$ is the minimizer of the
$n$th local objective, $f_n(\cdot)$.

The three different metrics $\norm{\nafs}$, ${\norm{\bx_*-\bxl}}$, and
${f(\bx_*)-f(\bxl)}$ can be related to one another.  Clearly, if all
agents have the same objective, so that $\bxl=\bx_*$, all three
metrics are zero.  Since ${\nabla f(\bxl)=\bzero}$, \eqref{eq:SCdef1a}
and \eqref{eq:SCdef1b} give directly
\begin{align}
  \frac{1}{L} \norm{\nafs} \le \norm{\bx_*-\bxl} \le \frac{1}{\mu} \norm{\nafs}.   \label{eq:bxlbnd}
\end{align}
Next, using Eq.~(2.1.10) in \cite[Th.~2.1.5]{nesterov2018lectures} and
Eq.~(2.1.24) in \cite[Th.~2.1.10]{nesterov2018lectures}, along with
${\nabla f(\bxl)=\bzero}$, gives
\begin{align}
  & \frac{1}{2L}\snorm{\nafs} 
  \le  f(\bx_*) -f(\bxl)  
  \le \frac{1}{2\mu} \snorm{\nafs}.
\end{align}

These inequalities show that if one of the heterogeneity metrics is
small (or large), the others must also be small (or large).  Note
that, further upper-bounding ${f(\bx_*)-f(\bxl) \leq f(\bzero)-f(\bxl)
}$, with $\bzero$ interpreted as a common initialization, yields the
data heterogeneity metric used in \cite{yuan2016convergence,8786146}
[${D^2/(2L)}$ therein].

\subsection{Multiple Consensus Rounds per Iteration}\label{sec:consstep}

Multiple consensus averaging rounds can be performed before the local
gradient update(s) are added, as suggested, for example in
\cite{8528465,koloskova2020unified}. With $K$ consensus rounds using
the same $\bW$, the equivalent iteration becomes \eqref{eq:unifiter}
with $\bW$ replaced by $\bW^K$.  The second-largest eigenvalue of
$\bW^K$ becomes ${\max\{\lambda_2^K,\lambda_N^K\}}$, correspondingly
affecting the convergence properties.

More generally, suppose that $K$ consensus rounds are performed, with
$\bW^k$ as weight matrix in round $k$. Suppose, furthermore, that the
agents store the consensus results after each round,
$\{\bW^k\bx_{t}\}$, and form a weighted average thereof, with
non-negative weights $\{ \alpha_k \}$ that satisfy ${\sum_{k=1}^K
  \alpha_k = 1}$. This results in the effective weight matrix
\begin{align}
  \bW' = \sum_{k=1}^K \alpha_k \bW^k.  \label{eq:Wprim}
\end{align}

As suggested for consensus problems in \cite{berthier2020accelerated},
$\{ \alpha_k \}$ may be optimized. For example, one may minimize ${
  \sum_{k=1}^K\alpha_k \max\{\lambda_2^k,\lambda_N^k\} }$ with respect
to $\{ \alpha_k \}$.  Alternatively, $\{ \alpha_k \}$ could be chosen
based on Chebyshev polynomials, as suggested by
\cite{scaman2017optimal} -- this is possible even if the eigenvalues
of $\bW$ are partly unknown.  The number of consensus rounds, $K$, may
be selected by considering the relative costs of communication and
computation.

\subsection{Breakdown of DGD for Bipartite Topologies}\label{sec:requirementsW}

All analysis so far has assumed that \eqref{eq:Weigs} holds, and specifically that
${\lambda_N>-1}$.  

For DGD, convergence breaks down if $\lambda_N=-1$. In this case,
\eqref{eq:eta2} cannot be satisfied as the numerator is zero.  The
case of $\lambda_N=-1$ can occur if the topology is bipartite.  (For a
bipartite graph, the eigenvalue spectrum is symmetric around $0$, so
${ \lambda_N = -\lambda_1 = -1 }$.)  For an example of a bipartite
topology where the DGD iteration \eqref{eq:stdDGD} diverges, take
$N=2$, $f_1(x)=f_2(x)=x^2/2$, and
\begin{align}
  \bW=\begin{bmatrix} 0 & 1  \\ 1 & 0 \end{bmatrix}.   \label{eq:exampleW}
\end{align}
Then  $\lambda_1=1$ and $\lambda_2=-1$.      The matrix that appears in
\eqref{eq:DGDfix2},
\begin{align}
  \bW-\eta \bA_{\bx\by} = \begin{bmatrix} -\eta & 1 \\ 1 & -\eta \end{bmatrix}
\end{align}
has the eigenvalues $\pm 1 -\eta$, so $\phi(\cdot)$ is not a
contraction for any $\eta$.  In fact, with ${\bx_0=[1 \ -1]^\T}$,
${\norm{\bx_t}\to\infty}$ as $t\to\infty$ in \eqref{eq:stdDGD}.  A
simple way to make DGD work for bipartite topologies is to tune the
consensus speed (see
Section~\ref{sec:noisygrad}.\ref{subsec:commnoise}.\ref{subsec:tuningspeed}):
provided that $\gamma<1$, we have ${ 1 - \gamma +\gamma\lambda_N >-1 }$ even if
$\lambda_N=-1$.

For diffusion, \eqref{eq:Weigs} can be relaxed to permit
$\lambda_N=-1$.  With $\lambda_N=-1$, the diffusion iteration
\eqref{eq:diffusionatc} converges; note, however, that $\bW^t$ does
not converge.

\subsection{Sufficient Conditions for $\bW$ to Satisfy \eqref{eq:Weigs}} \label{sec:suffcond}

A set of \emph{sufficient} conditions for \eqref{eq:Weigs} to hold, in
addition to the connectedness, $\bW^\T=\bW$, and ${\bW\bone=\bone}$
conditions, is that the elements of $\bW$ are non-negative and that
$\bW$ is not bipartite.  To see this, note that under these
conditions, by the Perron-Frobenius theorem
\cite[Th.~8.4.4]{matrixanalysis} there is a dominant eigenvector whose
elements are strictly positive, and whose corresponding eigenvalue has
multiplicity one and equals $\mnorm{\bW}$.  Since the eigenvectors of
$\bW$ are mutually orthogonal, all eigenvectors but the dominant one
have both positive and negative components.  Therefore, the dominant
eigenvector must be $\bone$, with eigenvalue $\lambda_1=1$.  Since
this eigenvector has multiplicity one, and the eigenvalues are
real-valued, $\lambda_2<\lambda_1$. There could be more eigenvalues
whose magnitudes equal $\mnorm{\bW}$, but the only possibility is
$-1$.  Since $\bW$ is not bipartite, it is primitive; then by
\cite[Th.~8.5.3]{matrixanalysis}, $\lambda_N>-1$.  Hence,
\eqref{eq:Weigs} is satisfied.

\subsection{Equal Edge Weights}

Additional insights into how the connectivity topology affects
convergence can be gained by starting with an undirected, unweighted
network and constructing $\bW$ as follows:
\begin{align}
  \bW = \bI - \epsilon \bL, \label{eq:avcons}
\end{align}
where $\bL$ is the Laplacian of the network and $\epsilon$ is a
positive constant.  Then ${\bW^\T=\bW}$, ${\bW\bone=\bone}$, and all
nonzero edge weights (excluding the loops) are equal to
$-\epsilon$. Let $\{\Lambda_n\}$ be the eigenvalues of $\bL$. Then
\begin{align}
  \lambda_n & = 1-\epsilon \Lambda_n, \\[-2mm]
  0 & =  \Lambda_1 < \Lambda_2 \le \cdots \le \Lambda_N \lea 2\kmax, 
\end{align}
where $\kmax$ is the largest node degree and (a) follows from the
Gershgorin disc theorem
\cite[Th.~6.1.1]{matrixanalysis}.\footnote{\setstretch{0.7}A tighter
inequality can be obtained, but is not of concern here.}  Hence, if
${\epsilon<1/\kmax}$, the eigenvalues of $\bW$ in \eqref{eq:avcons}
satisfy \eqref{eq:Weigs}.

Table~\ref{tab:eigs} shows $1-\lambda_2$ for equal edge weights and
some toy topologies.  The better connected the graph is (the smaller
the diameter), the better is the convergence behavior.

\begin{table}[t!]
  \centering
  \begin{tabular}{|Sc|Sc|}
    \hline
    \textbf{Topology} &\adjustbox{valign=c}{   $1-\lambda_2$}   \\
    \hline 
    Complete graph  &    $N\epsilon$  \\
    \hline
    Star &     $\epsilon$    \\
    \hline
    Line (path) & $ \displaystyle 4  \sin^2\pp{\dfrac{\pi}{2N}} \epsilon \approx \frac{\pi^2}{N^2}\epsilon$ \\
    \hline
    Ring (closed path) & $\displaystyle  4  \sin^2\pp{\dfrac{\pi}{N}} \epsilon \approx \frac{4\pi^2}{N^2}\epsilon $  \\
    \hline
  \end{tabular}
  \vspace*{3mm}
  \caption{$1-\lambda_2$ for some toy topologies with  equal edge  weights.}
  \vspace*{-5mm}
  \label{tab:eigs}
\end{table}

\section{Concluding Remarks}\label{sec:concl}

A rigorous analysis of the DGD and diffusion algorithms for
differentiable, strongly convex and smooth objectives and symmetric
weight matrices is advantageously approached by using our proposed
MHT+contraction framework.  Specifically, this independently yields,
first, a tight contraction to a fixed point ($\hat\bx$), and second, a
tight bound on the distance from $\hat\bx$ to the optimum ($\bx_*$).
The resulting analyses extend directly to multiple local gradient
updates, time-varying step sizes, sampling noise in the gradients,
noisy communication links, and random topologies.  In the noisy case,
the approach avoids assumptions on bounded gradient noise variance.

Our analyses unify, considerably simplify, and extend results in the
literature that were obtained using more complicated proof techniques.
This makes our framework an attractive way of describing, using, and
teaching decentralized convex optimization.  Also, many of the
presented results can be used, in a plug-and-play fashion, by
practitioners who want to establish convergence of their algorithms
without reproducing intricate technical arguments.

{The limitations of the MHT+contraction framework are: (1) the
  objectives must be (once) differentiable, strongly convex, and
  smooth; and (2) the weight matrix must be symmetric -- except in the
  random-topology case, where it is enough that this matrix is
  symmetric \emph{in expectation}.  We stress that if the strong
  convexity assumption were dropped, the theory of contraction
  mappings would no longer apply.  It remains an open problem whether
  our framework could be extended to handle deterministic asymmetric
  weight matrices.}

{Our bounds -- similar to results in the literature -- have
  the general form ${c^t+\varepsilon}$, where $c$ is a geometric
  factor and $\varepsilon$ is an asymptotic error. We showed that the
  geometric factors ($c$) obtained in our analysis are the best
  possible.  The bounds on the asymptotic error ($\varepsilon$)
  compare favorably with other known bounds (e.g.,
  \cite{yuan2016convergence,8786146}), but they could be improved upon
  further.}

We believe that our proposed framework may be used to analyze other
iterative algorithms for optimization over networks, such as gradient
tracking.  Another direction to explore is whether algorithms for
optimization over nonlinear manifolds (e.g.,
\cite{chen2021decentralized}) could be cast in our framework.

\onecolumn
\appendix

\subsection{An Auxiliary Result on Distance Bounding}\label{subsubsec:auxbnd}

Consider the solution with respect to $\bd$ of the linear equation
\begin{align}
  \pp{  \bZ\bA + \frac{1}{\eta} (\bI-\bW)} \bd = - \bZ\bv,  \label{eq:maineq}
\end{align}
where $\bZ$ is defined in \eqref{eq:defZ}, $ L \bI \succcurlyeq \bA
\succcurlyeq \mu \bI$, $\bW$ satisfies the requirements set forth in
Section~\ref{sec:DGD}.\ref{subsec:topology}, and the vector $\bv$
satisfies $\Piu\bv=\bzero$.  Then
\begin{align}
  \norm{\bd} & \le  \eta\frac{L}{\mu}\Lambda\norm{\bv},  \label{eq:distb}
\end{align}
where $\Lambda$ is defined in \eqref{eq:defLambda},
and where for diffusion we also require ${\eta\leq 1/(L\Lambda)}$.

\subsubsection*{Proof of \eqref{eq:distb}}
Multiplying \eqref{eq:maineq} from the left by $\Piu$ and using
$\Piu\bIW=\bzero$, in either case, yields $\Piu \bA \bd = \bzero$.
Then
\begin{align}
  \bzero =   \Piu \bA \pp{\Piu \bd +   \Piup \bd }= \frac{\bone^\T\bA\bone}{N} \Piu \bd +  \Piu \bA \Piup \bd
  = \frac{\bone^\T\bA\bone}{N} (\bd-\Piup \bd) +  \Piu \bA \Piup \bd,
\end{align}
which shows that
\begin{align}
  \bd=\pp{\bI - \frac{N}{\bone^\T\bA\bone} \Piu\bA}\Piup\bd=\pp{\bA^{-1} - \frac{N}{\bone^\T\bA\bone} \Piu}\bA\Piup\bd.  \label{eq:dpiupd}
\end{align}
Note that the matrix inside the second parenthesis in
\eqref{eq:dpiupd} satisfies
\begin{align}
  \frac{1}{\mu}\bI  \succcurlyeq  \bA^{-1} - \frac{N}{\bone^\T\bA\bone} \Piu \succcurlyeq \bzero, \label{eq:Apiu}
\end{align}
where the second term has only one nonzero eigenvalue, which is at
most $1/\mu$ because $\bone^\T\bA\bone \ge N\mu$.  Then
\eqref{eq:dpiupd} implies
\begin{align}
  \norm{\bd} \le \frac{L}{\mu} \norm{\Piup\bd}.  \label{eq:dnormLmu}
\end{align}
If $\Piup\bd=\bzero$ there is nothing to show; hence we assume that
$\Piup\bd\neq\bzero$.

\begin{itemize}
  \item For DGD ($\bZ=\bI)$, multiply \eqref{eq:maineq} from the left
    by $\bd^\T\Piup$ and insert \eqref{eq:dpiupd}; using
    $(\bI-\bW)\Piu=\bzero$ this gives
    \begin{align}
      \bd^\T\Piup \ppb{ \bA\pp{\bA^{-1} - \frac{N}{\bone^\T\bA\bone} \Piu}\bA  + \frac{1}{\eta} \bIW} \Piup\bd =     - \bd^\T\Piup\bv . \label{eq:dPipv1}
    \end{align}
    Using \eqref{eq:Apiu} and the fact that 
   {${\bI-\bW \succcurlyeq (1-\lambda_2)\Piup}$}, along with Cauchy-Schwarz inequality,
    \eqref{eq:dPipv1} implies that
    \begin{align}
      \frac{1-\lambda_2}{\eta} \snorm{\Piup\bd} & \le \norm{  \Piup\bd} \cdot \norm{\bv}, 
    \end{align}
    that is,
    \begin{align}
      \norm{  \Piup\bd} & \le  \frac{\eta}{1-\lambda_2} \norm{\bv}  .  \label{eq:normPupbd}
    \end{align}
    Equation \eqref{eq:normPupbd}, combined with \eqref{eq:dnormLmu},
    gives the expression for DGD in \eqref{eq:distb}.
 
  \item For diffusion ($\bZ=\bW$), instead multiply \eqref{eq:maineq}
    from the left by $\bd^\T\Piup\bIWp$. Using that
    ${\bIWp(\bI-\bW)=\Piup}$ {(since $\lambda_2<1$)}, this gives
    \begin{align}
      \bd^\T\Piup\bIWp\bW\bA\bd + \frac{1}{\eta}\bd^\T\Piup\bd  = -  \bd^\T\Piup\bIWp\bW\bv . \label{eq:bdpiupwv1}
    \end{align}
    Applying Cauchy-Schwarz inequality to \eqref{eq:bdpiupwv1} gives
    \begin{align}
      \frac{1}{\eta}\snorm{\Piup\bd} \le  \norm{\Piup\bd} \cdot \norm{\bIWp\bW\bv}  + 
      \norm{\Piup\bd} \cdot \norm{\bIWp\bW\bA\bd} .
      \label{eq:bdpiupwv2}
    \end{align}
    Dividing both sides of \eqref{eq:bdpiupwv1} by $\norm{\Piup\bd}$
    and inserting \eqref{eq:dpiupd} yields
    \begin{align}
      \frac{1}{\eta}\norm{\Piup\bd} & \le \norm{\bIWp\bW\bv}  + \norm{\bIWp\bW\bA  \pp{\bA^{-1} - \frac{N}{\bone^\T\bA\bone} \Piu}  \bA  \Piup\bd} \\
      & \le \norm{\bIWp \bW\bv} +
      \mnorm{\bIWp\bW} \mnorm{\bA   \pp{\bA^{-1} - \frac{N}{\bone^\T\bA\bone}\Piu} \bA  }  \norm{ \Piup\bd}
      \\
      & \lea \norm{\bIWp \bW\bv} +
      L\mnorm{\bIWp\bW}  \norm{ \Piup\bd},
    \end{align}
    where in (a) we used that \eqref{eq:Apiu} implies $ \bA
    \pp{\bA^{-1} - \frac{N}{\bone^\T\bA\bone}\Piu} \bA \preccurlyeq
    L\bI $.  Hence,
    \begin{align}
      \norm{\Piup\bd} & \le \frac{\eta}{1- \eta L\mnorm{\bIWp\bW}} \norm{\bIWp \bW\bv} \lea 2\eta \norm{\bIWp \bW\bv} \le 2\eta \mnorm{\bIWp \bW}\norm{\bv},
    \end{align}
    where (a) holds if $\eta \le 1/(L \Lambda
    )$.\footnote{\setstretch{0.7}This bound can be tightened slightly
    by enforcing a stricter condition on $\eta$. For an arbitrary
    ${\alpha>1}$, we have ${1/(1- \eta L\mnorm{\bIWp\bW})\le\alpha}$
    if ${ \eta\le (\alpha-1)/(\alpha L\mnorm{\bIWp\bW}) }$. We took
    $\alpha=2$, which results in the cleanest-looking, but not
    tightest possible, bound.}  Using again \eqref{eq:dnormLmu} we get
    the formula for diffusion in \eqref{eq:distb}. $
    \hfill\blacksquare$

\end{itemize}

\subsection{Derivation of \eqref{eq:dgdfinal} and \eqref{eq:diffusionfinal}} \label{app:multilocal}

The strategy is to obtain a linear equation that relates
$\bxhs$ to $\nafs$, where $\hat\bx$ is the fixed point of
$\varphi(\bx)$, and then from this equation upper-bound
$\norm{\bxhs}$.

We first show the following auxiliary equality:
\begin{align}
  (\bI-\eta\bA_{T-2})\cdots(\bI-\eta\bA_i) =  \bI - \eta \sum_{k=i}^{T-2} \bA_k + \bcO(\eta^2T^2), \qquad i=0,\ldots,T-2,  \label{eq:AOetaT2}
\end{align}
where $\{\bA_i\}$ is a sequence of matrices such that
$\mu\bI\preccurlyeq \bA_i\preccurlyeq L\bI$.  To establish
\eqref{eq:AOetaT2}, define, for $i=0,\ldots,T-2$,
\begin{align}
  \bE_i =  (\bI-\eta\bA_{T-2})\cdots(\bI-\eta\bA_i) - \pp{ \bI - \eta \sum_{k=i}^{T-2} \bA_k}. \label{eq:Eidef}
\end{align}
Clearly, \eqref{eq:AOetaT2} holds if
\begin{align}
  \mnorm{\bE_i}\le \eta^2 (T-2-i)^2 L^2, \label{eq:Einorm}
\end{align}
for $i=0,\ldots,T-2$.  Equation \eqref{eq:Einorm} holds for $i=T-2$,
since ${\bE_{T-2}=\bzero}$.  Suppose, as induction hypothesis, that
\eqref{eq:Einorm} is true for some $i$. Then
\begin{align}
  \mnorm{\bE_{i-1}} & = \mnorm{(\bI-\eta\bA_{T-2})\cdots(\bI-\eta\bA_{i-1}) - \pp{ \bI - \eta \sum_{k=i-1}^{T-2} \bA_k} } \\
  & = \mnorm{\ppb{ \bE_i + \pp{ \bI - \eta \sum_{k=i}^{T-2} \bA_k} } (\bI-\eta\bA_{i-1}) - \pp{ \bI - \eta \sum_{k=i-1}^{T-2} \bA_k} } \\
  & = \mnorm{ \bE_i(\bI-\eta\bA_{i-1}) + \eta^2 \sum_{k=i}^{T-2} \bA_k \bA_{i-1}} 
  \lea \mnorm{ \bE_i } + \eta^2\sum_{k=i}^{T-2}\mnorm{ \bA_k \bA_{i-1}} \\
  & \le \eta^2(T-2-i)^2 L^2 + \eta^2 (T-1-i) L^2 \le \eta^2(T-1-i)^2 L^2,
  \label{eq:inductionIAT2}
\end{align}
where in (a) we used that ${\mnorm{ \bI-\eta\bA_{i-1} } \le 1}$.
Hence, \eqref{eq:Einorm} holds for $i-1$.  By induction over $i$,
\eqref{eq:Einorm}, and therefore \eqref{eq:AOetaT2}, hold for all $i$.

Now, to relate $\bxhs$ to $\nafs$, first introduce the
following auxiliary variables ${\tilde\bx_0,\ldots,\tilde\bx_T}$ to
describe the mapping $\varphi(\bx)$ ($\tilde\bx_i$ represents the
result after the $i$th local gradient update):
\begin{align}
  \tilde\bx_0 & = \bx, \\
  \tilde\bx_{i+1} & = \phi_i(\tilde\bx_i) = \tilde\bx_i - \eta\nabla f(\tilde\bx_i) , \quad  i=0,\ldots,T-2 \\
  \tilde\bx_T & = \varphi(\bx)  = \phi_T(\tilde\bx_{T-1}) = \bW \tilde\bx_{T-1} - \eta\bZ\nabla f(\tilde\bx_{T-1}),  \label{eq:localiterA}
\end{align}
where $\bZ$ is defined in \eqref{eq:defZ}. 
Note that when $\varphi(\bx)$ is applied to its fixed point,
$\hat\bx$, it holds that
\begin{align}
  \tilde\bx_0=\tilde\bx_T =   \varphi(\hat\bx)=\hat\bx.
\end{align}
From \eqref{eq:localiterA}, we obtain
\begin{align}
  \tilde \bx_0 - \bx_* & = \hat \bx - \bx_* , \\
  \tilde \bx_{i+1} - \bx_* & = \tilde\bx_i - \eta\nabla f(\tilde\bx_i) - \bx_* \\
  & \eqa \tilde\bx_i  -\eta[\nafs + \bA_i (\tilde\bx_i-\bx_*)] - \bx_* \\
  & = (\bI-\eta\bA_i)(\tilde\bx_i-\bx_*) - \eta \nafs , \qquad i=0,\ldots,T-2,
\end{align}
where in (a) we applied the MHT to $f(\cdot)$ at the points $\bx_*$ and
$\tilde\bx_i$, and used $\bA_i$ as a shorthand for the kernel $\bA_{\bx_*\tilde\bx_i}$.
By induction over $i$, and using \eqref{eq:AOetaT2},
\begin{align}
   \tilde \bx_{T-1} - \bx_* 
  & =  (\bI-\eta\bA_{T-2})\cdots (\bI-\eta\bA_0)(\tilde\bx_0-\bx_*) 
  - \eta { \sum_{i=1}^{T-1} (\bI-\eta\bA_{T-2})\cdots(\bI-\eta\bA_i)    } \nafs \\
  & = \pp{ \bI - \eta\sum_{i=0}^{T-2} \bA_i + \bcO(\eta^2T^2)}(\tilde\bx_0-\bx_*) 
  -  \eta \pp{ (T-1)  \bI  - \eta \sum_{i=1}^{T-1}  \sum_{j=i}^{T-2}  \bA_j + \bcO(\eta^2T^3)  } \nafs.\label{eq:xT1xstar1}
\end{align}
Also, 
\begin{align}
\hat\bx - \bx_*   & = \tilde \bx_T - \bx_* \\
  & = \bW \tilde\bx_{T-1} - \eta \bZ \nabla f(\tilde\bx_{T-1}) - \bx_* \\
         & = \bW (\tilde\bx_{T-1}-\bx_*) - \eta \bZ \nabla f(\tilde\bx_{T-1}) \\
         & \eqa \bW (\tilde\bx_{T-1}-\bx_*) - \eta \bZ [\nafs + \bA_{T-1}(\tilde\bx_{T-1}-\bx_*)] \\
  & = (\bW - \eta\bZ\bA_{T-1})(\tilde\bx_{T-1}-\bx_*) - \eta \bZ \nafs \\
  & \eqb (\bW - \eta\bZ\bA_{T-1})\ppb{ \pp{ \bI - \eta\sum_{i=0}^{T-2} \bA_i + \bcO(\eta^2T^2)}(\tilde\bx_0-\bx_*) 
    - \eta\pp{(T-1)\bI -  \eta \sum_{i=1}^{T-1}  \sum_{j=i}^{T-2}  \bA_j + \bcO(\eta^2T^3)  } \nafs }
  \\ & \qquad - \eta \bZ \nafs,
   \label{eq:xtxstar1}
\end{align}
where in (a) we again used the MHT on $f(\cdot)$, and in (b) we used
\eqref{eq:xT1xstar1}.  Finally, collecting the terms in
\eqref{eq:xtxstar1} gives
\begin{align}
& \qquad \pp{ \bI - \bW + \eta\bW \sum_{i=0}^{T-2} \bA_i + \eta\bZ\bA_{T-1} + \bcO(\eta^2T^2)}(\hat \bx - \bx_*) 
\\ & = \pp{ -\eta  \ppb{(T-1)\bW+\bZ} + \eta^2 (T-1)\bZ\bA_{T-1}+\eta^2\bW \sum_{i=1}^{T-1}  \sum_{j=i}^{T-2}  \bA_j + \bcO(\eta^3T^3) } \nafs
  .  \label{eq:maineq_xt}
\end{align}

We will now, based on \eqref{eq:maineq_xt}, obtain a bound on ${\norm{
    \bxhs} }$, by writing
\begin{align}
  \bxhs & =  \Piup (\bxhs) + \Piu (\bxhs)   \label{eq:splitortho1} 
\end{align}
and separately considering the two terms ${\norm{ \Pi^\perp_\bone
    (\bxhs)}}$ and ${\norm{ \Piu (\bxhs)} }$, and
using that
\begin{align}
  \norm{\bxhs} & \le  \norm{\Piup (\bxhs)} + \norm{\Piu (\bxhs)}.
  \label{eq:splitortho}
\end{align}

First, multiplication of \eqref{eq:maineq_xt} from the left by
$(\bI-\bW)^\dagger$, using the identity ${ (\bI-\bW)^\dagger
  (\bI-\bW)=\Piup }$, gives
\begin{align}
  &\qquad \ppb{ \Piup + \eta (\bI-\bW)^\dagger\pp{ \bW\sum_{i=0}^{T-2}\bA_i + \bZ\bA_{T-1}}   + \bcO(\eta^2T^2)} (\hat \bx - \bx_*) \\
  & = 
  \ppb{ -\eta(\bI-\bW)^\dagger \ppb{(T-1)\bW+\bZ} +\eta^2(\bI-\bW)^\dagger\pp{(T-1) \bZ\bA_{T-1}+\bW \sum_{i=1}^{T-1}  \sum_{j=i}^{T-2}  \bA_j} + \bcO(\eta^3T^3)}
\nafs.  \label{eq:maineq_xt_piperp}
\end{align}
By taking the norm of both sides of \eqref{eq:maineq_xt_piperp} and
using the triangle inequality, we obtain an upper bound on the first
term in \eqref{eq:splitortho}:
\begin{align}
  \norm{ \Piup (\hat \bx - \bx_*) }  & \le 
  \eta \mnorm{(\bI-\bW)^\dagger \ppb{(T-1)\bW+\bZ}} \norm{\nafs} + \O(\eta T) \norm{\hat \bx - \bx_*} + \O(\eta^2T^2).
  \label{eq:maineq_xt_piperp_n}
\end{align}

Next, multiplication of \eqref{eq:maineq_xt} from the left by
$\bone^\T$ and using again \eqref{eq:splitortho1}, along with
\eqref{eq:bonenafs}, and that ${\bW\bone=\bZ\bone=\bone}$ gives
\begin{align}
  \bone^\T   \pp{ \sum_{i=0}^{T-1} \bA_i  + \bcO(\eta T^2) }\Piu(\hat \bx - \bx_*) 
  & = - \bone^\T \pp{ \sum_{i=0}^{T-1} \bA_i  + \bcO(\eta T^2) } \Piup ( \hat \bx - \bx_*)
  + \bone^\T \ppb{  \eta\sum_{i=1}^{T-1}  \sum_{j=i}^{T-1}  \bA_j  +\bcO(\eta^2T^3) } \nafs .
  \label{eq:maineq_xt_pi}
\end{align}
By using the properties of $\{\bA_i\}$ we establish that
\begin{align}
  (\mu+\O(\eta T))\norm{\Piu(\hat \bx - \bx_*)}
  & = \frac{1}{\sqrt{N}} (\mu+\O(\eta T))\left|\bone^\T(\hat \bx - \bx_*)\right|  \\
  & \le \frac{1}{\sqrt{N}} \left| \pp{ \frac{1}{TN}\sum_{i=0}^{T-1} \bone^\T\bA_i \bone + \O(\eta T)} \bone^\T(\hat \bx - \bx_*) \right| \\
  & = \frac{1}{T\sqrt{N}} \left| \bone^\T\pp{\sum_{i=0}^{T-1} \bA_i  + \bcO(\eta T^2) }\Piu(\hat \bx - \bx_*) \right| \\
  & = \frac{1}{T\sqrt{N}} \left| - \bone^\T\pp{ \sum_{i=0}^{T-1} \bA_i  + \bcO(\eta T^2) } \Piup ( \hat \bx - \bx_*)
    +  \bone^\T\pp{ \eta\sum_{i=1}^{T-1}  \sum_{j=i}^{T-1}   \bA_j    +\bcO(\eta^2T^3) } \nafs \right| \\
  & \le \pp{L + \O(\eta T)} \norm{\Piup ( \hat \bx - \bx_*)} + \frac{1}{T}\ppb{ \eta L\frac{T(T-1)}{2} + \O(\eta^2T^3)} \norm{ \nafs },
  \label{eq:maineq_xt_pinorm}
\end{align}
 which gives the following upper bound on the second term in
 \eqref{eq:splitortho}:\footnote{\setstretch{0.7}Note that if
 $\ppb{1+\O(\eta)} x = a + \eta b + \O(\eta^2)$ then $x=(1+\O(\eta)) a
 + \eta b + \O(\eta^2)$.}
\begin{align}
  \norm{\Piu(\hat \bx - \bx_*)}
  & \le \pp{\frac{L}{\mu}+\O(\eta T)} \norm{ \Piup(\hat \bx - \bx_*) } +  \eta \frac{(T-1)L}{2\mu}\norm{\nafs } + \O(\eta^2T^2) \\
  & \le \eta \frac{L}{\mu} \pp{ \mnorm{(\bI-\bW)^\dagger \ppb{(T-1)\bW+\bZ}} \norm{\nafs} + \frac{T-1}{2}\norm{\nafs }}
  \\ & \qquad  + \O(\eta T)\norm{\hat \bx - \bx_* } + \O(\eta^2 T^2). 
  \label{eq:maineq_xt_pinorm2}
\end{align}

Adding \eqref{eq:maineq_xt_piperp_n} and \eqref{eq:maineq_xt_pinorm2},
using \eqref{eq:splitortho}, gives
\begin{align}
  \norm{\hat \bx - \bx_*}
  & \le \eta \ppb{ \frac{L}{2\mu}(T-1) + \pp{1+\frac{L}{\mu}} \mnorm{(\bI-\bW)^\dagger \ppb{(T-1)\bW+\bZ}}} \norm{\nafs}
  + \O(\eta T)\norm{\hat \bx - \bx_* } + \O(\eta^2 T^2), 
  \label{eq:maineq_xt_normfinal}
\end{align}
from which, finally,
\begin{align}
  \norm{\hat \bx - \bx_*}
  & \le \eta \ppb{ \frac{L}{2\mu}(T-1) + \pp{1+\frac{L}{\mu}} \mnorm{(\bI-\bW)^\dagger \ppb{(T-1)\bW+\bZ}}} \norm{\nafs}
  + \O(\eta^2 T^2).   \label{eq:maineq_xt_normfinal2}
\end{align}

\subsubsection*{Explicit Bounds without $\O(\eta^2 T^2)$ Terms}

Throughout the treatment of multiple local gradient updates, we give
bounds on $\norm{\hat\bx-\bx_*}$ that contain higher-order terms
expressed in terms of $\O(\cdot)$.  This choice was made only to
maximize simplicity of the exposition, and the bounds with $\O(\cdot)$
are sufficient to show how $\norm{\hat\bx-\bx_*}$ scales with
different parameters of the problem.

But importantly, using the bound on $\mnorm{\bE_i}$ provided in
\eqref{eq:Einorm}, every $\O(\cdot)$-term may be replaced by an
explicit upper bound.  In particular, let $\alpha>1$ be arbitrary. If
${g(\eta T) \le h(\eta T) + \O(\eta^2T^2)}$ then there exists a
constant $\eta_0T_0$ such that ${ g(\eta T) \le \alpha h(\eta T) }$
when $\eta T\ge \eta_0 T_0$. While explicit expressions for $\eta_0T_0$
can be obtained, they become rather intricate and do not add insight.

\subsubsection*{Additional Remark}\label{subapp:remark1}

The following observation, which follows from the calculation above,
will be used in the analysis of the noisy case.  By taking the norm on
both sides of \eqref{eq:xT1xstar1}, we obtain $\norm{\tilde
  \bx_{T-1}-\bx_*} = \norm{\bxhs} + \O(\eta T)$ and by a
slight amendment of the calculation (re-doing the recursion up to $i$
instead of $T-1$) that
\begin{align}
   \norm{\tilde \bx_{i}-\bx_*}  = \norm{\bxhs} + \O(\eta T), \qquad i=0,\ldots, T-1.
\end{align}
Then for $i=0,\ldots, T-1$,
\begin{align}
  \norm{ \tilde \bx_i } \le \norm{\bx_*} + \norm{ \tilde\bx_i - \bx_*} 
  \le  \norm{\bx_*} + \norm{ \hat\bx - \bx_*} +\O(\eta T) \eqa  \norm{\bx_*} + \O(\eta T),  \label{eq:unibndxihat}
\end{align}
where (a) follows from \eqref{eq:maineq_xt_normfinal2}.

\subsection{Gradient Noise in Stochastic GD (and DGD/Diffusion) Can Have Unbounded Variance}\label{sec:unboundednoise}

To see why gradient noise in stochastic GD (and DGD) may have unbounded
variance, consider GD with linear regression in two
dimensions. Suppose the parameter is the slope, $x$, of a straight
line that passes through the origin, and suppose three data points are
available: $(1,1)$, $(1,-1)$, and $(2,0)$.  The objective is
\begin{align}
  f(x) =  \frac{1}{3} \ppb{ (1-x)^2+ (-1-x)^2+ (0-2x)^2},
\end{align}
each term corresponding to each respective data point.  For $f(x)$, we
have $f'(x)=4x$ and $\mu=L=4$, so with $\eta=1/4$, GD on
$f(x)$ would yield the optimal solution, $x=0$, in a single step.

Now suppose stochastic GD is applied, sampling one of
the data points (objectives) uniformly at random, and computing the
derivative (gradient).  This results in the noisy gradient,
\begin{align}
  &    \begin{dcases} 
         \frac{d}{dx}  {  (1-x)^2}  = 2x-2, & \mbox{with probability $1/3$} \\
         \frac{d}{dx}  {  (-1-x)^2 } = 2x + 2, & \mbox{with probability $1/3$} \\
         \frac{d}{dx}  {  (0-2x)^2 } = 8 x, & \mbox{with probability $1/3$} 
       \end{dcases} \\
  & = 4x     + \epsilon, 
\end{align}
where $\epsilon$ is the gradient noise. This noise has the
distribution (conditional on $x$)
\begin{align}
  \epsilon  = 
  \begin{cases}
    -2x-2  , & \mbox{with probability $1/3$} \\
    2-2x   , & \mbox{with probability $1/3$} \\
    4x , & \mbox{with probability $1/3$} 
  \end{cases},
\end{align}
and the  moments
\begin{align}
  \E{\epsilon | x} & = 0, \\
  \E{\epsilon^2 | x } & = \frac{8}{3} + 8x^2 . \label{eq:varianceLR}
\end{align}
The second moment in \eqref{eq:varianceLR} has the form in
\eqref{eq:altvarcond}, and is unbounded.

Linear-regression examples in arbitrary dimensions, and with arbitrary
numbers of data points, can be similarly constructed.

An example of a model that results in bounded gradient noise variance
is when the objective is sampled from a family of functions whose
gradients are uniformly bounded.\footnote{\setstretch{0.7}For differentiable objective
functions, the gradient is uniformly bounded if and only if the
function is Lipschitz continuous.  Such Lipschitz continuity of the
objective should not be conflated with $L$-smoothness, which is
equivalent to Lipschitz-continuity of the \emph{gradient}.}  However,
this model is incompatible with the assumption on strong convexity of
the objective, since strongly convex functions cannot have uniformly
bounded gradients.

An example of a model with a strongly convex objective that results in
bounded gradient noise variance is when the objective is sampled from
a family of convex (but not strictly convex) functions with uniformly
bounded gradients, to which a deterministic, strongly convex
regularization term is added.

\end{document}